\documentstyle[preprint,tighten,aps,eqsecnum,epsfig,floats,12pt]{revtex}
\def\bc{\begin{center}}
\def\ec{\end{center}}
\def\be{\begin{eqnarray}}
\def\ee{\end{eqnarray}}

\def\d#1#2{\frac{\displaystyle #1}{\displaystyle #2}}
\newcommand{\omits}[1]{}

\def\diff{di\hspace{-0.2em}f\hspace{-0.3em}f}

\def\r{\partial}
\def\Dl{\Delta}

\def\Si{\Sigma}
\def\Om{\Omega}

\def\dl{\delta}
\def\eps{\epsilon}
\def\ka{\kappa}
\def\th{\theta}
\def\om{\omega}

\def\del{\nabla}

\newcommand{\ul}{\underline}

\newcommand{\tT}{\tilde T}
\newcommand{\tR}{\tilde R}

\newcommand{\eqo}{\stackrel{\circ}{=}}

\newcommand{\tD}{\tilde D}

\newcommand{\bD}{\bar D}
\newcommand\btd{\raise 2pt
\hbox{$\hat\bigtriangledown$}\hskip 1.5pt}
\newcommand\bt{\raise 2pt
\hbox{$\bigtriangledown$}\hskip 1.5pt}

\def\NPB{{\it Nucl. Phys.}~{\bf B}}
\def\PRD{{\it Phys. Rev.}~{\bf D}}
\def\PRL{{\it Phys. Rev. Lett }}

\def\CQG{{\it Class. Quantum Gravity }}
\def\CMP{{\it Commun. Math. Phys. }}

\def\CTP{{\it Commun. Theor. Phys. }}

\begin{document}
\hfill April 8, 2003

\vfill
\bc{\Large \bf On Diffeomorphism Invariance and Black Hole Entropy} \ec

\bigskip
\begin{center}
Chao-Guang Huang$^{a,c}$\footnote{Email: huangcg@mail.ihep.ac.cn},\quad
Han-Ying Guo$^{b,c}$\footnote{Email: hyguo@itp.ac.cn}, \quad
Xiaoning Wu$^{d}$\footnote{Email: wuxiaon@yahoo.com.cn} 
\end{center}
\begin{center}{\small
$^a$ Institute of High Energy Physics, Chinese
Academy of
Sciences,\\
\quad P.O. Box 918(4), Beijing 100039, China.\\
$^b$ CCAST, World Laboratory, B.O. Box 8730, Beijing 100080, China\\
$^c$ Institute of Theoretical Physics, Chinese Academy of Sciences,\\
\quad  P.O. Box 2735, Beijing 100080,
China.\\
$^d$ Max-Plank-Insitut f\"ur Gravitationsphysik,
Am M\"uhlenberg 1, 14476 Golm, Germany.
}
\end{center}
\hspace{2cm}

\vfill

\bc \parbox{13cm} {{\bf Abstract.}\quad  The Noether-charge 
realization and the Hamiltonian
realization for the $\diff({\cal M})$ algebra in diffeomorphism
invariant gravitational theories are studied in a covariant
formalism.  For the Killing vector fields, the Nother-charge
realization leads to the mass formula as an entire vanishing
Noether charge for the vacuum black hole spacetimes in general
relativity and the corresponding first law of the black hole
mechanics. It is analyzed in which sense the Hamiltonian
functionals form the $\diff({\cal M})$ algebra under the Poisson
bracket and shown how the Noether charges with respect to the
diffeomorphism generated by vector fields and their variations in
general relativity form this algebra.  The asymptotic behaviors of 
vector fields
generating diffeomorphism of the manifold with boundaries are
discussed.  In order to get more precise estimation for the
``central extension" of the algebra, it is analyzed in the
Newman-Penrose formalism and shown that the ``central extension"
for a large class of vector fields is always zero on the Killing
horizon. It is also checked whether the Virasoro algebra may be
picked up by choosing the vector fields near the horizon. The
conclusion is unfortunately negative.} \ec

\vspace{1cm} \bc
\parbox{10cm}
\ec

\vfill
\newpage
\tableofcontents

\newpage
\section{Introduction}

 It is well known that the relation between the
symmetry and related conservation law is one of
the cornerstones in physics. For the spacetime
manifolds in general relativity (GR), the
diffeomorphisms form an infinite dimensional
group under the composition. Therefore, it should
also play certain important roles.

In the canonical formalism, Brown and Henneaux \cite{bh} showed
that the asymptotic symmetry group for the three-dimensional
gravity with negative cosmological constant is the
pseudo-conformal group in 2-dimension.  Its canonical generators
form a pair of Virasoro algebras with nontrivial central charges.
Strominger \cite{str} suggested that with the help of the Cardy
formula \cite{cd}, the asymptotic symmetry and the central charge
can be used to explain the statistical origin of the
Bekenstein-Hawking entropy \cite{BH} of the 3-dimensional black
hole. In 1999, Carlip attempted to generalize the
Brown-Henneaux-Strominger construction to the black holes in any
dimension \cite{c1} but met some conceptual problems \cite{ph}.
Recently, the generalization is analyzed more carefully by Park in
the canonical formalism \cite{park}.

Carlip \cite{{c2},{c3}} also tries to realize the Strominger's
idea in any dimension by using the covariant phase-space formalism
developed by Wald {\it et al} \cite{lw}-\cite{iw2}.  Carlip's
approach has been generalized to the charged black hole
\cite{jy1}, the dilaton black hole \cite{jy2}, the Kaluza-Klein
black hole \cite{jy3}, the quantum correction \cite{c4}, {\it
etc}.  In the meanwhile, Dreyer, Ghosh and Wi\'sniewski pointed
out that this covariant phase-space formalism by Carlip contains
technical flaws \cite{dgw}.  Koga related the flaws to the
singular behavior of the asymptotic forms of the relevant
quantities \cite{K}.  In the study of the asymptotic symmetries on
the Killing horizon in spherically symmetric spacetimes, Koga also
showed that the algebra of the Poisson brackets does not acquire
nontrivial central charges \cite{K}. An alternative version of
this approach is also studied in the first-order gravity formalism
\cite{A}.

In the covariant phase-space formalism
\cite{{c2},{c3}}, the Hamiltonian functional
conjugate to a vector field is treated as the
generator of diffeomorphism algebra just like the
Hamiltonian in the canonical approach. 
The diffeomorphism algebra is assumed to be
realized by the Poisson bracket or by Dirac
bracket \cite{Db} on the constraint surface.
Unfortunately, the Hamiltonian functional conjugate
to a vector field $\xi^a$, as was pointed out by
Wald \cite{{wald},{iw}}, does not always exist for a given
boundary condition.  In addition, in the lack of
the definition of the Poisson bracket and thus
the Dirac bracket in the covariant phase-space
formalism, Carlip borrowed the Poisson bracket and
the Dirac bracket from the ADM formalism
\cite{Db}--\cite{T}.  Although 
there are a number
of studies on the definition and properties of
the Poisson brackets in the covariant phase-space
approach \cite{fpr}, such as the ones in the de Donder-Weyl
field theory, it is still no explicit proof
for their equivalence with the Poisson bracket in
the ADM formalism. In fact, in the ADM formalism
the Poisson bracket is defined on the basis of
the 1+3 decomposition of a spacetime manifold,
but in the covariant
phase-space formalism there is no explicit
splitting of the time and the space components.

On the other hand, however, in both classical and
quantum field theories, if
a Lagrangian possesses certain symmetries, such
as gauge symmetry and Poincar\'{e} symmetry, the
corresponding Lie algebras can 
be generated by the Noether charges of the
conservation currents with respect to the
symmetries. This general approach should also be
available for the diffeomorphism invariance of
the diffeomorphism-invariant theories of gravity such
as GR.

The main purposes of the present paper are four folds.  First, it
is to obtain, in a covariant formalism, the Noether charge by the
horizontal variation of the Lagrangian for diffeomorphism
invariant gravitational theories.  It is shown that for the
Killing vector fields the covariant formalism leads to the vacuum
black hole mass formula as a vanishing Noether charge as well as
the first law of the black hole mechanics in vacuum GR. Secondly,
it is analyzed in which sense the Hamiltonian functionals form the
$\diff({\cal M})$ algebra in the covariant formalism. It is also
shown how the $\diff$(${\cal M}$) algebra with possible ``central
extension" can be realized by virtue of the Noether charges in
vacuum GR. The third purpose is to show why the ``central
extension" of the algebra vanishes on shell near the horizon. When
the boundary of the (partial) Cauchy surface is fixed on the
horizon, it is obvious that the ``central extension" vanishes
because the realizations themselves are trivial. In the approach
\cite{{c2},{c3}}, the ``central extensions" of the $\diff$(${\cal
M})$ and the Virasoro algebra are dependent on the asymptotic
behaviors of the vector fields specially chosen that correspond to
the situation that the boundary of the partial Cauchy surface is
not fixed.  In order to estimate more precisely the asymptotic
behaviors of such a kind of the vector fields, the Newman-Penrose
(N-P) formalism is employed in this paper. The advantage of the
N-P formalism is obvious.  It not only can be set up on the
horizon as well as the spacetime region with $r>r_h$, but also may
avoid the arbitrariness in the asymptotic behavior estimation.
It is then shown that the ``central extension" of $\diff$(${\cal
M})$ for the diffeomorphism generated by the vector field
$\xi^a=Tl^a+Rn^a$ with $R \sim O(\chi_K^{})$ is always zero on the
horizon. The fourth purpose of the present paper is to check
whether the Virasoro algebra can be picked up by choosing the
vector fields near the horizon.  It is shown in the N-P formalism
that algebra of the vector fields selected in \cite{c2} under the
Lie bracket does not give the Virasoro algebra.  Even in the sense
of \cite{c2} that the Lie brackets of the vector fields form a
Virasoro algebra on the horizon, the algebra of the Hamiltonian
functionals conjugate to the vector fields is trivial, too, and
should not be regarded as a Virasoro algebra.  This indicates that
such a kind of pure symmetry analysis is not enough to give the 
statistical or conformal field theory origin of the black hole
entropy.

The paper is organized as follows:  In the next section, we review
the covariant formalism including the Noether currents with
respect to the diffeomorphisms generated by vector fields and
their charges in a diffeomorphism-invariant theory.  We also show
how the vacuum black hole mass formula as a whole is a total
vanishing Noether charge with respect to the combination of the
Killing vector fields for the stationary axisymmetric black holes
in GR and re-derive the fist law of the black hole mechanics for
this configuration \cite{wghw}.   In section III, the Hamiltonian
realization and the Noether-charge realization of the
diffeomorphism algebra in the covariant phase-space formalism 
are studied. In
section IV, the diffeomorphism algebra for a part of a manifold is
discussed.  Sec. V is devoted to the asymptotic behavior of the
vector fields generating the diffeomorphism.  In section VI, the
``central extension" of $\diff(\cal M)$ is estimated in the N-P
formalism and shown why it always vanishes near the horizon. In
section VII, we show in what sense the Virasoro algebra can be
obtained as a subalgebra of diffeomorphism algebra and why it is
trivial.  Finally, we summarize this paper and make some
discussion in section VIII.   The correspondence, in the N-P formalism, of the asymptotic conditions on the horizon
proposed by Carlip
\cite{{c2},{c3}} is given in Appendix A.  In Appendix B, some 
relations used in the calculation in the N-P formalism are listed.

\section{Noether Charges in
Diffeomorphism Invariant Gravitational Theory \label{s2}}

 Let us consider a
diffeomorphism-invariant gravitational theory on
4-dimensional spacetime manifold $\cal M$ \cite{note1}.

\subsection{Noether currents and their charges with respect to diffeomorphism}

The horizontal variation of the Lagrangian 4-form
${\bf L}$ of such a kind of theories induced by a vector field
$\xi^a$ can be written as 
\cite{{wald},{iw},{wghw}}
\begin{eqnarray} \label{s2.f1}
\hat \dl_\xi {\bf L}={\bf E} \hat
\dl_\xi g+ d{\mbox {\boldmath $\Theta$}}(g,
\hat\dl_\xi g),
\ee
where ${\bf E}=0$ gives rise to the Euler-Lagrange
equation for the theory and ${\mbox {\boldmath
$\Theta$}}(g,\hat \dl_\xi g)$ is the symplectic
potential 3-form.  On the other hand, using the
Lie derivative ${\cal L}_\xi$, one has
\be \label{s2.f2}
\hat\dl_\xi{\bf L} ={\cal L}_\xi{\bf L}=d(\xi
\cdot {\bf L}).
\end{eqnarray}
Equating Eqs.(\ref{s2.f1}) and (\ref{s2.f2}), one gets
\begin{eqnarray}
\label{s2.Neq}
d*{\bf j}(\xi) +{\bf E}\hat \dl_\xi g=0,
\end{eqnarray}
where $*$ is the Hodge star and
\be
\label{s2.cc1}
{\bf j}(\xi) = *({\mbox {\boldmath $\Theta$}}
(g,\hat \dl_\xi g) -\xi \cdot {\bf L})
\ee
is the Noether current 1-form
with respect to the diffeomorphism generated by the
given vector field $\xi^a$. Its (entire) Noether
charge is given by the integral over a generic Cauchy
surface $\Si \in {\cal M}$
\be
\label{s2.NC1}
Q(\xi)=\int_{\Sigma} *{\bf j}(\xi).
\ee

In order to obtain the differential formula for mass, i.e. the
first law of the black hole mechanics, it is needed the variation
relation of this Noether current. To this end, it is natural to
calculate the following bi-variation with respect to vertical and
horizontal variation $\delta$ and $\hat\delta$ of ${\bf L}$,
\begin{eqnarray}
 \delta \hat\delta_\xi {\bf L} = \delta({\bf
E}\hat\delta_\xi g+ d {\bf \Theta} (\hat\delta_\xi g))
=\delta d(\xi\cdot{\bf L})\,.
\end{eqnarray}
Therefore, from these equations, it follows the
variation relation of this Noether current
\begin{eqnarray}
0=\delta({\bf E} \hat \dl_\xi g)+ d{\delta *\bf
j}\,,
\end{eqnarray}
where we have used the commutative property
between the vertical variation operator and the
differential operator, i.e. $\delta d=d\delta$.
Thus, for the conservation of  variation of the
Noether current (\ref{s2.cc1}), i.e.
\begin{eqnarray}
d{\delta *\bf j}=0\,,
\end{eqnarray}
the necessary and sufficient condition is
\begin{eqnarray}\label{s2.ddg}
\delta({\bf E}\hat\dl_\xi g)=0\,.
\end{eqnarray}

\subsection{Noether charges for the Killing vectors and the vacuum mass\\
formula in GR }

In vacuum GR, the Lagrangian 4-form in
units of $G=c=1$ reads
\be
{\bf L} = \d 1 {16\pi}
R {\mbox {\boldmath $\eps$}},
\ee
where $R$ is
the scalar curvature and ${\mbox {\boldmath
$\eps$}}$ the volume 4-form.  The symplectic
potential takes the form of
\be
{\mbox {\boldmath
$\Theta$}}_{abc} (g, \hat\dl_{\xi}g) = \d 1
{16\pi}[\del^d (g_{ef} \hat\dl_\xi
g^{ef})-\del_e \hat\dl_\xi g^{de}] {\mbox
{\boldmath $\eps$}}_{dabc}. \ee
Thus, the Noether
current and its charge may be explicitly written
as
\be \label{s2.cc2}
{\bf j}_{a}(\xi) =
\frac{1}{8\pi}G_{ab}\xi^b -\frac{1}{16\pi} \nabla
^b (\nabla _b \xi_a-\nabla _a \xi_b) \ee and \be
\label{NC2} Q(\xi) = \d 1 {8\pi}\int _\Si *
(G_{ab} \xi^b) - \frac{1}{16 \pi}\int _{\r \Si}
*d{\mbox {\boldmath $\xi$}}, \ee
respectively,
where $G_{ab}$ is the Einstein tensor and $\r
\Si$  the boundary of the Cauchy surface.

In particular, for the stationary axisymmetric 
black holes in the vacuum GR and the Killing
vector
\be \label{s2.kv} \chi_K ^a = t_K^a + \Om_H
\phi_K^a, \ee
$Q(\chi_K^{\ })$ vanishes and leads to the black hole mass formula
\cite{wghw}, where $t_K^a$ and $\phi_K^a$ are the timelike and
spacelike Killing vectors of the spacetime, respectively, $\Om_H$
the angular velocity on the horizon and hereafter, the subscript
`$K$' denotes the vector (or the corresponding one-form) being the
Killing one. The reasons for the vanishing Noether charge leading
to the black hole mass formula are as follows:  For a Killing
vector,
\be
\frac{1}{2}(\nabla_b\nabla^a\chi_K^b-\nabla_b\nabla^b\chi_K^a)
= R^a_b\chi_K^b, \ee
which is zero on shell and
leads to
\be {\bf j}(\chi_K^{\ })=-\frac{1}{16\pi}R\xi_a +
\frac{1}{4\pi}R_{ab}\xi^b
=-\frac{1}{16\pi}R\xi_a-\frac{1}{8\pi} \nabla ^b
(\nabla _b \xi_a-\nabla _a \xi_b) .\ee
Consequently, $Q(\chi_K^{\ })$ vanishes on shell.
By the definition of the mass and the angular
momentum of the black hole, \begin{eqnarray}
{\cal Q}_\infty(\chi_K^{\
})=-\frac{1}{8\pi}\int_{S_{\infty}}*d{\mbox
{\boldmath $\chi$}}_K^{\ }
 = M - 2 \Omega_HJ.
\end{eqnarray}
On the other hand,
\begin{eqnarray}
{\cal Q}_{S_H}(\chi_K^{\ }) = - \frac{1}{8\pi}\int_{S_H}*d{\mbox
{\boldmath $\chi$}}_K^{\ } =\frac{\kappa}{4\pi}A,
\end{eqnarray}
where $\kappa$ is the surface gravity and $A$ the
area of the cross section of the event horizon.
${\cal Q}$ will be referred to as the partial Noether
charge of a close surface.
For the whole asymptotically flat region, the
Cauchy surface emanates from the bifurcation
surface and extends to the spatial infinity.
Thus, the boundary of the Cauchy surface
$\partial \Sigma$ should be $S_H^{(-)}\cup
S_\infty$, and the Noether charges should take
the form
\be \label{s2.NC2o}
Q(\chi)=Q(\chi_K^\infty) + Q(\chi_K^H) = {\cal
Q}_{\infty}(\chi_K^{\ })-{\cal Q}_{S_H}(\chi_K^{\ }), \ee
where $Q(\chi_K^\infty)$ and $Q(\chi_K^H)$ are the Noether
charges for the Cauchy surface with compact interior and the
one with compact infinity, respectively.
This directly leads to the vacuum mass formula 
in GR as a vanishing (entire) Noether charge.
Namely,
\begin{eqnarray}\label{s2.mformula}
\frac{\kappa}{4\pi}A- M + 2 \Omega_H J =0 .
\end{eqnarray}

 It should be mentioned
that there are somewhat differences between the present
Noether charge approach 
and the one by Wald {\it et al}.  First,
in their approach the Noether charge for a spacetime
manifold with the Cauchy surface possessing both
interior and infinite asymptotically flat
boundaries has not been considered rather what
the Cauchy surface has been introduced is the one
with compact infinity. Secondly, the orientation
of the bifurcation surface $S_H$ is taken as
positive rather than the one induced from the
Cauchy surface with compact infinity. In fact, if
the induced orientation could be taken, the
entropy were no longer the Noether charge even in
their approach but the negative one.

The Noether charge may be defined for a finite
region of a spacetime.  When $\Sigma$ is not chosen to be
the whole of the Cauchy surface but its portion
with two boundaries $B_1$ and $B_2$ such that $A_{B_2}>A_{B1}$,
where $A_{B}$ stands for the area of the surface $B$,
\be
\label{s2.NC3}
Q(\xi) = {\cal Q}_{B_2}(\xi) - {\cal Q}_{B_1}(\xi)
\ee
gives the Noether charge for the portion of the
spacetime region $R\times\Si_p$.

\subsection{The first law of black hole
mechanics in vacuum GR}

Let us now derive, from the variation relation of this vanishing
Noether charge, the differential formula for mass, i.e.\ the first
law of black hole mechanics, for the vacuum stationary
axisymmetric black holes in GR by the variational approach.

For the vacuum gravitational fields in GR,  the
variation relation of this vanishing Noether
current Eq. (\ref{s2.ddg}) becomes
\begin{eqnarray}
0=\frac{1}{16\pi}\delta({G_{ab}}\hat\dl_{\chi^{}_K}
g^{ab})+ \delta[{\nabla^a {\bf j}_a}(\chi^{}_K)]\,,
\end{eqnarray}
where ${\bf j}_a(\chi^{}_K)$ is given by Eq.~(\ref{s2.cc2}). As
was mentioned before, it is obvious that the
conserved current $\bf j$ vanishes for the
Killing vector field $\chi^a_K$ (\ref{s2.kv})  if the
vacuum Einstein equation holds.  Further, if the
variation or the perturbation $\delta g$ is
restricted in such a way that both $g$ and
$g+\delta g$ are stationary axisymmetric black
hole configurations and $\chi^a_K$ is the Killing
vector (\ref{s2.kv}), the vanishing Noether current
and the variation of the dual of the
conserved current should also vanish, i.e.
$\delta {*\bf j}=0$ as well.

Thus it is straightforward to get
\begin{eqnarray}\label{s2.BCH}
0=\frac{1}{8\pi}\int_{\Sigma}
\delta[R_{ab}\chi^b_K{d \sigma}^a] =\delta
[M-\frac{\kappa}{4\pi}A-2\Omega_H J ],
\end{eqnarray}
and  Eq.~(\ref{s2.ddg}) is also satisfied. Here
$\Sigma$ is a 
Cauchy surface.

It should be noticed that  Eq.~(\ref{s2.BCH}) is
just the start point of Bardeen, Carter and
Hawking's calculation for the first law of the
black hole mechanics \cite{BCH}.  
As was required in \cite{BCH}, under the above
perturbations, the positions of
event horizon and the two Killing vector fields
in Eq.~(\ref{s2.kv}) are unchanged. Consequently, as
long as following what had been done in
\cite{BCH},  Eq.~(\ref{s2.BCH}) definitely 
leads to the differential formula for mass, i.e., the first law of
black hole mechanics, among the stationary
axisymmetric black hole configurations
\begin{eqnarray}\label{s2.1law}
\delta M-\frac{\kappa}{8\pi}\delta
A-\Omega_H\delta J=0\,.
\end{eqnarray}

Thus both the mass formula (\ref{s2.mformula}) and its
differential formula (\ref{s2.1law}), i.e., the first law of black
hole mechanics for the stationary axisymmetric black hole
configurations are all derived from the diffeomorphism invariance
of the Lagrangian by the Noether charge via variational approach.
Especially, they are in certain sense the vanishing Noether charge
and its perturbation among the stationary axisymmetric black hole
configurations.

\section{On Realizations of $\diff$(${\cal M}$) Algebra
\label{s3}}

In this section, we shall consider two
different ways of realizations for the algebra
$\diff$(${\cal M}$), i.e. the Hamiltonian
realization and the Noether charge realization.
\

\subsection{The Hamiltonian realization}
 It has been shown \cite{BHS} that in
gauge theories the flux of the symplectic current
\be \label{s3.sc}
\int _\Sigma *[\delta_2 {\bf j}(\phi,\delta_1 \phi) -
\delta_1 {\bf j}(\phi,\delta_2 \phi)]
\ee
is equivalent to the Hamiltonian bracket in the
reduced phase space. In Eq.(\ref{s3.sc}) ${\bf j}$
is the Noether current, which is the dual of the
symplectic potential ${\mbox {\boldmath
$\Theta$}}(\phi, \delta \phi) =\pi_\phi \delta \phi$
and $\delta_2 {\bf j}(\phi,\delta_1 \phi) -
\delta_1 {\bf j}(\phi,\delta_2 \phi)$ is the
(pre)symplectic current, where $\delta$ is a  vertical
variation.

The attempts to generalize the equivalence to the diffeomorphism
invariant theories have been made \cite{lw}-\cite{iw}\cite{Torre}.
The (pre)symplectic current in a diffeomorphism invariant theory and
its flux over a spacelike hypersurface are
\be    \label{s3.om1}
{\mbox {\boldmath$\om$}}(\phi, \delta_1 \phi, \delta_2 \phi) =
\delta_1 {\mbox {\boldmath $\Theta$}}(\phi, \delta_2\phi) -
\delta_2  {\mbox {\boldmath $\Theta$}}(\phi, \delta_1\phi)
\ee
and
\be \label{s3.Om1a}
\Om(\phi, \delta_1 \phi, \delta_2 \phi)=\int _\Si
{\mbox {\boldmath$\om$}}(\phi, \delta_1 \phi, \delta_2 \phi),
\ee
respectively.   In Eqs.(\ref{s3.om1}) and (\ref{s3.Om1a}), $\dl_1$
and $\dl_2$ have been generalized to arbitrary variations, which
include both vertical and horizontal variations.   The
(pre-)symplectic structure, {\it i.e.} the flux of the (pre)symplectic
current Eq.(\ref{s3.Om1a}), has a remarkable property
\cite{{lw},{Torre}}
\be \label{s3.Om10}
\Om(\phi, \delta_1 \phi, \hat \delta_\xi \phi)=
\Om(\phi, \delta_1 \phi, {\cal L}_\xi \phi) =0.
\ee
Namely, if one of the two variations is horizontal one, then the
flux of the (pre)symplectic current must vanish.
In \cite{wald} it has been
shown explicitly that when the variation $\dl_2$ in Eq.(\ref{s3.Om1a}) is
restricted to the horizontal variation $\hat \dl _\xi$,
\be
\label{s3.dH0}
\delta H(\xi) = \Om (\phi, \delta \phi, \hat \dl_\xi \phi),
\ee
where
\be \label{s3.HQr}
H(\xi) = Q(\xi) - \int _{\partial \Si} \xi \cdot {\bf B},
\ee
${\bf B}$ is the 3-form satisfying
\be
\label{s3.B}
\delta \int _{\partial \Sigma} \xi \cdot {\bf B}(\phi) =
\int _{\partial \Si}\xi\cdot {\mbox {\boldmath $\Theta$}}(\phi, \delta \phi).
\ee
It should be noted that in Eqs.(\ref{s3.dH0})--(\ref{s3.B})
\be \label{s3.vcond}
\delta \xi =0
\ee
is required \cite{wald}.  In general, the vertical variations satisfy
the condition (\ref{s3.vcond}), but the horizontal ones do not.

In \cite{c2} the variation $\dl$ in Eq.(\ref{s3.dH0}) is further
restricted to the horizontal variation $\hat \dl_{\xi_2}={\cal L}_{\xi_2}$
so that the Cartan formula ${\cal L}_{\xi} = d \circ i_\xi + i_\xi \circ d$
for any differential form can be used,
and the identification of the flux of the presymplectic current and the Poisson bracket  $\{\ , \ \}$  is made.  Then, Eq.(\ref{s3.dH0})
becomes
\be
\label{s3.dH}
\hat\delta _{\xi_2} H(\xi_1) = \Om (\phi, \hat\dl_{\xi_2} \phi, \hat\dl_{\xi_1} \phi)
=\{ H(\xi_1), H(\xi_2) \}.
\ee
However, there are two difficulties in Eq.(\ref{s3.dH}).  First,
in accordance with the requirement (\ref{s3.vcond}) for the first equality,
we should have $\hat \dl_{\xi_2}\xi_1=0$, which is obvious in contradiction
with the basic formula ${\cal L}_{\xi_2} \xi_1 = [\xi_2, \xi_1]$.  Second,
it is the non-trivial flux of symplectic current that can be used to define
the Poisson bracket.  Unfortunately, the property (\ref{s3.Om10}) says that the flux
of presymplectic current  $\Om (\phi, \hat\dl_{\xi_2} \phi, \hat\dl_{\xi_1} \phi)$
vanishes, so that the second equality is not valid.
In order to reconcile the first contradiction, we have two choices: 

{\bf A}. The two vector fields are not arbitrary
but fall into special classes. For example, the
vector fields coincide up to a multiplication
constant at the boundaries of Cauchy surfaces if
$\hat\delta_\xi H$ is the algebraic summation of
boundary terms.

{\bf B}. Since the first variations of fields are
linear and homogeneous functions of $\xi$, we may
define a new quantity by
\be  \label{s3.dlF} \bar \dl _{\xi_2} F(\xi_1):=
\hat\dl _{\xi_2} F(\xi_1) -
F(\hat\dl_{\xi_2}\xi_1) =\hat\dl _{\xi_2}
F(\xi_1) -F([\xi_2, \xi_1]). \ee
By use of Eqs.(\ref{s3.dlF}), (\ref{s2.f1}), (\ref{s2.cc1}), one gets
\be
\bar\dl_{\xi_2}*{\bf j}(\xi_1) 
&=& \bar\dl_{\xi_2} {\mbox {\boldmath $\Theta$}}(\xi_1)
- \xi_1 \cdot \bar\dl_{\xi_2}{\bf L} \nonumber \\
&=& \bar\dl_{\xi_2} {\mbox {\boldmath $\Theta$}}(\xi_1)
- \xi_1 \cdot [{\bf E} \bar\dl_{\xi_2}g +
d{\mbox {\boldmath $\Theta$}}(\bar\dl_{\xi_2}g)] \nonumber \\
&=& \bar\dl_{\xi_2} {\mbox {\boldmath $\Theta$}}(\xi_1)
-{\cal L}_{\xi_1} {\mbox {\boldmath $\Theta$}}(\bar\dl_{\xi_2}g)
+ d[\xi_1 \cdot {\mbox {\boldmath $\Theta$}}(\bar\dl_{\xi_2}g)].
\ee
where on-shell condition has been used.  Therefore, if the
variation equation
\be  \label{s3.B'}
\bar \delta \int _{\partial \Sigma} \xi \cdot {\bf B}(\phi) =
\int _{\partial \Si}\xi\cdot {\mbox {\boldmath $\Theta$}}(\phi, \bar \delta \phi),
\ee
then the first equality of Eq.(\ref{s3.dH}) should be modified as
\be
\label{s3.dbarH}
\bar\delta _{\xi_2} H(\xi_1) = \Om' (\phi, \hat\dl_{\xi_2} \phi, \hat\dl_{\xi_1} \phi),
\ee
where
\be \label{s3.Om1b}
\Om'(\phi, \hat \dl_{\xi_1} \phi, \hat \dl_{\xi_2} \phi)&=&\int _\Si
{\mbox {\boldmath$\om$}}'(\phi, \hat \dl_{\xi_1} \phi, \hat \dl_{\xi_2} \phi) \nonumber \\
&=&  \int _\Si
\bar \dl_{\xi_1} {\mbox {\boldmath $\Theta$}}(\phi, \hat \dl_{\xi_2}\phi) -
\bar \dl_{\xi_2} {\mbox {\boldmath $\Theta$}}(\phi, \hat \dl_{\xi_1}\phi).
\ee
In Eq.(\ref{s3.Om1b}), ${\cal L}_{\xi_1} {\mbox {\boldmath $\Theta$}}(\bar\dl_{\xi_2}g)$
has been written as
$\bar \dl_{\xi_1} {\mbox {\boldmath $\Theta$}}(\phi, \hat \dl_{\xi_2}\phi)$
on account that the appearance of $\bar\dl_{\xi}$ implies that
$[\xi_1,\xi_2]$ is discarded in the calculation.

Even when the triviality problem of the flux of the presymplectic current
is put by, there is ambiguity in the identification of the Poisson bracket
with the flux of the presymplectic currents, taking the form of Eq.(\ref{s3.dH}) or the form like
\be
\Om'(\phi, \hat \dl_{\xi_1} \phi, \hat \dl_{\xi_2} \phi)= \{ H(\xi_1), H(\xi_2) \}.
\ee

According to the Brown-Henneaux analysis in the canonical approach \cite{bh},
the diffeomorphism algebra in terms of the Dirac bracket for the
boundary terms of the Hamiltonian functionals in the covariant phase space
formalism is, in Ref. \cite{c2}, represented in the form
\be
\label{s3.ha}
\{ J(\xi_1), J(\xi_2) \}^*= J([\xi_1, \xi_2])+K^H(\xi_1, \xi_2),
\ee
where $K^H(\xi_1, \xi_2)$ is the possible central extension in the
Hamiltonian realization of $\diff$($\cal M$).  The upper index $H$ stands
for this realization.

\subsection{The Noether-charge realization}

 In gauge theories the (pre)symplectic current (\ref{s3.om1})
and the antisymmetric combination of the vertical
variations of the Noether current
\be    \label{s3.om2a}
\delta_1 * {\bf j} (\phi, \delta_2\phi) - \delta_2 *{\bf j}(\phi, \delta_1\phi)
=: {\mbox{\boldmath $\tilde \om$}} (\phi,
\delta_1 \phi, \delta_2 \phi).
\ee
coincide.  In general, however, they are
different from each other in diffeomorphism
invariant theories.\omits{After replaced the
vertical variations $\dl$ by the horizontal ones
$\hat \dl_\xi$, Eq.(\ref{s3.om2a}) becomes
\be    \label{om2b}
{\mbox{\boldmath $\tilde \om$}} (\phi, \hat
\delta_1 \phi, \hat \delta_2 \phi) = \hat
\delta_1 * {\bf j} (\phi,\hat \delta_2\phi) -
\hat \delta_2 *{\bf j}(\phi,\hat \delta_1\phi).
\ee
}
With the help of the flux of the current (\ref{s3.om2a}), it is possible to realize
the $\diff$ algebra by use of Noether charges \cite{ghw}.

In order to set up the Noether-charge realization, let us
consider the two successive horizontal
variations of the Lagrangian 4-form induced by
two vector fields $\xi_1$ and $\xi_2$ denoted by $\hat\dl_{\xi_1}$ and
$\hat\dl_{\xi_2}$, respectively,
\be
\label{s3.2v}
\hat \dl_{\xi_1} \hat \dl_{\xi_2}{\bf
L} = \hat \dl_{\xi_1} [{\bf E}\hat \dl_{\xi_2} g
+ d {\mbox {\boldmath $\Theta$}}(g,\hat
\dl_{\xi_2 } g)]= \hat \dl_{\xi_1} d (\xi_2 \cdot
{\bf L}).
\ee
Exchanging the order of the variation, one obtains
\be
\label{s3.2v'}
\hat \dl_{\xi_2} \hat \dl_{\xi_1}{\bf
L} = \hat \dl_{\xi_2} [{\bf E}\hat \dl_{\xi_1} g
+ d {\mbox {\boldmath $\Theta$}}(g,\hat
\dl_{\xi_1 } g)]= \hat \dl_{\xi_2} d (\xi_1 \cdot
{\bf L}).
\ee
Subtraction of the two equations gives rise to
\be
\label{s3.2vs}
\hat \dl_{\xi_1} \hat \dl_{\xi_2}{\bf L}
 - \hat \dl_{\xi_2} \hat \dl_{\xi_1}{\bf L}
&=& \hat \dl_{\xi_1} [{\bf E}\hat \dl_{\xi_2} g
+ d {\mbox {\boldmath $\Theta$}}(g,\hat
\dl_{\xi_2 } g)] -\hat \dl_{\xi_1} [{\bf E}\hat \dl_{\xi_2} g
+ d {\mbox {\boldmath $\Theta$}}(g,\hat
\dl_{\xi_2 } g)] \nonumber \\
&=& \hat \dl_{\xi_1} d (\xi_2 \cdot
{\bf L})- \hat \dl_{\xi_2} d (\xi_1 \cdot
{\bf L}).
\ee
Since the Lie bracket of two vector fields gives
a new vector field, Eqs.(\ref{s2.f1}) and (\ref{s2.f2})
can also apply to the Lie bracket of the two vector fields,
namely,
\begin{eqnarray}
\label{s3.f3}
\hat \dl_{[\xi_1, \xi_2]} {\bf L}={\bf E} \hat
\dl_{[\xi_1, \xi_2]} g+ d{\mbox {\boldmath $\Theta$}}(g,
\hat\dl_{[\xi_1, \xi_2]} g) = d([\xi_1, \xi_2] \cdot {\bf L}).
\end{eqnarray}
Subtracting Eq.(\ref{s3.f3}) from Eq.(\ref{s3.2vs}) and using
the identity
\begin{eqnarray}
\label{s3.i}
[\hat \dl_{\xi_1}, \hat
\dl_{\xi_2}] = \hat \dl_{[\xi_1,
\xi_2]},
\end{eqnarray}
one has
\be
\label{s3.2vss}
0&=&[\hat \dl_{\xi_1}, \hat \dl_{\xi_2}]{\bf L}
 - \hat \dl_{[\xi_1, \xi_2]} {\bf L} \nonumber \\
&=& \hat \dl_{\xi_1} [{\bf E}\hat \dl_{\xi_2} g
+ d {\mbox {\boldmath $\Theta$}}(g,\hat
\dl_{\xi_2 } g)] -\hat \dl_{\xi_1} [{\bf E}\hat \dl_{\xi_2} g
+ d {\mbox {\boldmath $\Theta$}}(g,\hat
\dl_{\xi_2 } g)] -{\bf E} \hat
\dl_{[\xi_1, \xi_2]} g+ d{\mbox {\boldmath $\Theta$}}(g,
\hat\dl_{[\xi_1, \xi_2]} g)  \nonumber \\
&=& \hat \dl_{\xi_1} d (\xi_2 \cdot
{\bf L})- \hat \dl_{\xi_2} d (\xi_1 \cdot
{\bf L}) -d([\xi_1, \xi_2] \cdot {\bf L}) .
\ee
It follows that
\begin{eqnarray}
d\{\hat \dl_{\xi_1}[*{\bf j}(\xi_2)]-
\hat \dl_{\xi_2}[*{\bf j}(\xi_1)]
-*{\bf j}([\xi_1,\xi_2])\}
&=& \hat \dl_{\xi_2}({\bf
E}\hat \dl_{\xi_1} g) - \hat \dl_{\xi_1}({\bf
E}\hat \dl_{\xi_2} g)+{\bf E}\hat \dl_{[\xi_1,
\xi_2]} g.
\end{eqnarray}
Namely, the  combination of the current 1-forms as a Noether-like
current 1-form \be\label{s3.ccs} {\bf k}(\xi_1,\xi_2) =
* \hat \dl_{\xi_1}[*{\bf j}(\xi_2)] -
* \hat \dl_{\xi_1} [*{\bf j}(\xi_2)] -
{\bf j}([\xi_1,\xi_2])  
\ee
is conserved as long as
\be \label{s3.dEdg}
\hat \dl_{\xi_2}({\bf
E}\hat \dl_{\xi_1} g) - \hat \dl_{\xi_1}({\bf
E}\hat \dl_{\xi_2} g)+{\bf E}\hat \dl_{[\xi_1,
\xi_2]} g = 0.
\ee
Obviously, Eq.(\ref{s3.dEdg}) is valid on shell.  It should be
noted that the condition Eq.(\ref{s3.dEdg}) is weaker than the
on-shell condition. If the charge of ${\bf k}(\xi_1,\xi_2)$ is
denoted by $-K^Q(\xi_1, \xi_2)$ with the upper index $Q$ for the
Noether-charge realization, then $K^Q(\xi_1, \xi_2)$
satisfies\footnote{In the present paper, we consider the case that
the Cauchy surface keeps unchanged under the horizontal
variation.}
\be
\label{s3.preQb}
\hat \delta _{\xi_1} [Q(\xi_2)] - \hat \delta _{\xi_2} [Q(\xi_1)]
=Q([\xi_1,\xi_2]) - K^Q(\xi_1, \xi_2).
\ee

The left hand side of Eq.(\ref{s3.preQb}) is nothing but the flux of
the current (\ref{s3.om2a})
over a Cauchy surface with the substitution of the vertical variation by the
horizontal one
\be  \label{s3.Om2}
\tilde \Omega(\phi, \hat \delta_1 \phi, \hat \delta_2 \phi)&=&\int _\Si
{\mbox{\boldmath $\tilde \om$}}(\phi, \hat \delta_1 \phi, \hat \delta_2 \phi) \nonumber \\
&=& \hat \delta_1 Q(\phi,\hat \delta_2\phi) -
\hat \delta_2 Q(\phi,\hat \delta_1\phi).
\ee
Again, the requirement Eq.(\ref{s3.vcond}) should be satisfied
in the second variation in Eq.(\ref{s3.Om2}) in
accordance with Eq.(\ref{s3.om2a}), which results in either the vector
fields falling into specific classes or $\hat\dl$ being replaced
by $\bar\dl$ in the second variation.  \omits{For the first choice,
Eq.(\ref{s3.preQb}) is a trivial realization as Eq.(\ref{s3.ha}).}
For the second choice,
\omits{\ul{??How to derive these formulas directly from
Lagrange??}
}
by use of Eq. (\ref{s3.dlF}) the left-hand side of Eq.(\ref{s3.preQb}) becomes
\be
 && (\hat \dl _{\xi_1} Q)(\xi_2) +Q(\hat \dl _{\xi_1} \xi_2)-
(\hat \dl _{\xi_2} Q)(\xi_1) - Q(\hat \dl _{\xi_2} \xi_1) \nonumber \\
& = &(\hat \dl _{\xi_1} Q)(\xi_2) +Q([\xi_1, \xi_2])-
(\hat \dl _{\xi_2} Q)(\xi_1) - Q([\xi_2, \xi_1])
- Q([\xi_1, \xi_2]) \nonumber \\
& = &\bar \dl _{\xi_1} Q(\xi_2) -
\bar \dl _{\xi_2} Q(\xi_1) + 2 Q([\xi_1, \xi_2]).
\ee
Therefore, the Noether charges and their variations on shell
form the algebraic relation of $\diff$(${\cal M}$)
\be
\label{s3.Qb}
\bar \delta _{\xi_2} Q(\xi_1) - \bar \delta _{\xi_1} Q(\xi_2)
=Q([\xi_1,\xi_2]) + K^Q(\xi_1, \xi_2),
\ee
where $K^Q(\xi_1, \xi_2)$ is the possible central extension.

It should be stressed again that in Eq.(\ref{s3.Qb}) the variation of $Q$ is
carried out under the condition that the first vector $\xi$ keeps
unchanged.  In another word, the variation does not act on the vector
field.  The left-hand side of Eq.(\ref{s3.Qb}) is
the flux of the exterior variation of the dual of Noether current Eq.(\ref{s3.Om2})
after the replacement of $\hat\dl$ by $\bar\dl$.  It is remarkable that
the ambiguity in the definition of the Poisson bracket in the Hamiltonian realization
does not appear in the Noether-charge realization because the Poisson bracket
does not come in the Noether-charge realization at all.

\subsection{The 2-cocycle condition for the central extension in Noether-charge realization}

 If an algebra realization
satisfies the Jacobi identity, its central
extension always obeys 2-cocycle condition
from the Jacobi identity and vice versa.  Now, let us consider
\begin{eqnarray}
\label{s3.p3}
{\cal C}_{\{1,2,3\}}\hat \delta_{\xi_1}
\{\hat \delta_{\xi_2} [Q(\xi_3)] -
\hat \delta_{\xi_3} [Q(\xi_2)]\} = 
{\cal C}_{\{1,2,3\}} \hat \delta_{\xi_1} [Q([\xi_2,\xi_3])]-
{\cal C}_{\{1,2,3\}} \hat \delta_{\xi_1}[K^Q(\xi_2,\xi_3)],
\end{eqnarray}
where ${\cal C}_{\{1,2,3\}}$ denotes the circular summation.
By use of identity (\ref{s3.i}), the left-hand side of (\ref{s3.p3}) can be written as
\begin{eqnarray}
{\cal C}_{\{1,2,3\}}\{\hat \delta_{\xi_1}\hat \delta_{\xi_2}
[Q(\xi_3)] - \hat \delta_{\xi_2} \hat \delta_{\xi_1}[Q(\xi_3)]\}
= {\cal C}_{\{1,2,3\}}\hat \delta_{[\xi_1,\xi_2]} [Q(\xi_3)]
={\cal C}_{\{1,2,3\}}\hat \delta_{[\xi_2,\xi_3]} [Q(\xi_1)].
\end{eqnarray}
It follows from (\ref{s3.p3}) that
\begin{eqnarray}
\label{s3.K1}
{\cal C}_{\{1,2,3\}} \hat \delta_{\xi_1}[K^Q(\xi_2,\xi_3)]=
{\cal C}_{\{1,2,3\}}\{ \hat \delta_{\xi_1} [Q([\xi_2,\xi_3])]-
\hat \delta_{[\xi_2,\xi_3]}[Q(\xi_1)]\}.
\end{eqnarray}
On the other hand, from (\ref{s3.preQb}) one has
\begin{eqnarray}
\label{s3.K2}
\hat \delta_{[\xi_2,\xi_3]} [Q(\xi_1)] -
\hat \delta_{\xi_1} [Q([\xi_2,\xi_3])] = Q([[\xi_2,\xi_3],\xi_1])
-K^Q([\xi_2,\xi_3],\xi_1).
\end{eqnarray}
Eqs. (\ref{s3.K1}) and (\ref{s3.K2}) and the Jacobi identity
\begin{eqnarray}
\label{s3.jcb}
{\cal C}_{\{1,2,3\}}[\hat\delta_{\xi_1},[\hat\delta_{\xi_2},
\hat\delta_{\xi_3}]] =0
\end{eqnarray}
give rise to
\begin{eqnarray}
\label{s3.KK}
{\cal C}_{\{1,2,3\}} \hat \delta_{\xi_1}[K^Q(\xi_2,\xi_3)] =
{\cal C}_{\{1,2,3\}} K^Q([\xi_1,\xi_2],\xi_3).
\end{eqnarray}
On account of identity (\ref{s3.i}), the Jacobi identity
becomes
\begin{eqnarray}
\label{s3.vjcb}
{\cal C}_{\{1,2,3\}}[\hat\delta_{\xi_1}, \hat\delta_{[\xi_2,\xi_3]}] =0.
\end{eqnarray}
Applying it on {\bf L}, one has
\be
{\cal C}_{\{1,2,3\}}\{ \hat \delta_{\xi_1} [Q([\xi_2,\xi_3])]-
\hat \delta_{[\xi_2,\xi_3]}[Q(\xi_1)]\} =0,
\ee
which leads to
\be
{\cal C}_{\{1,2,3\}} \hat \delta_{\xi_1}[K^Q(\xi_2,\xi_3)] =
{\cal C}_{\{1,2,3\}} K^Q([\xi_1,\xi_2],\xi_3) = 0.
\ee
This is the 2-cocycle condition for the central
extension.

\section{Diffeomorphism Algebra for a Portion of a Manifold \label{s4}}

\begin{figure}[htbp]
\centerline{\epsfbox{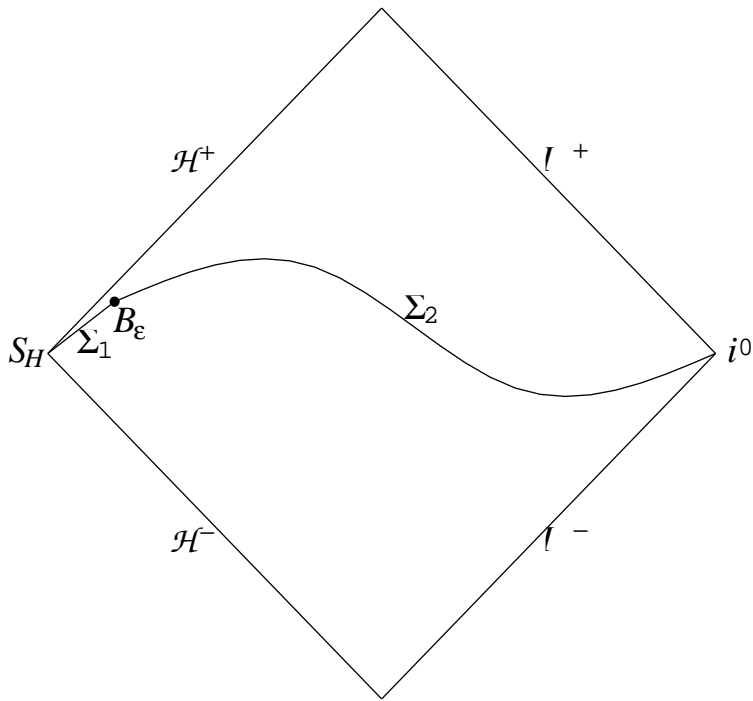}}
\label{cpd}
\caption{The Carter-Penrose diagram of an asymptotically-flat
region for stationary axisymmetric spacetimes.  ${\cal I}^+$,
${\cal I}^-$, and $i^0$ are the future and past null infinity,
and the spatial
infinity, respectively.  ${\cal H}^+$, ${\cal H}^-$, and
$S_H$ are the future and past event horizon, and the bifurcation
surface, respectively.  $\Sigma=\Sigma_1 \cup \Sigma_2$ is
the Cauchy surface for the whole of the asymptotically flat region.
$\Sigma_1$ and $\Sigma_2$ match at $B_\epsilon$.  $\eps$ tends
to $0$ so that $\Sigma_1$ tends to the event horizon.}
\end{figure}

In this section, we focus on the case of stationary,
asymptotically flat, axisymmetric
spacetimes in vacuum GR.  The Carter-Penrose diagram for the
asymptotically flat region of a stationary
axisymmetric spacetime is shown in the figure 1.

Now, we consider the Cauchy surface which is
combined by two pieces $\Si_1$ and $\Si_2$ as
shown in the figure 1.  Namely,
\begin{eqnarray}
\label{s4.sC}
\Sigma=\Sigma_1 \cup \Sigma_2.
\end{eqnarray}
$\Sigma_1$ emanates from the bifurcation surface
$S_H$, extends almost along the generator of the
event horizon and ends at the certain place of the
stretched Killing horizon \cite{c2} denoted by
$B_\epsilon$.  At the end of calculation, $\Si_1$
tends to the event horizon by taking $\epsilon
\rightarrow 0$. $\Sigma_2$ is a spacelike
hypersurface matching $\Si_1$ at $B_\epsilon$ and
extending to the spatial infinity. The boundaries of
$\Si_1$ and $\Si_2$ are $\partial
\Sigma_1=S_H^{(-)} \cup B_\epsilon$ and $\partial
\Sigma_2=B^{(-)}_\epsilon \cup S_\infty$,
respectively. Such a choice of the Cauchy
surface, may be called a combined Cauchy surface,  will
not affect the fact that the Noether charge
corresponding to the Killing vector field
(\ref{s2.kv}) for the whole asymptotically flat
region leads to the black hole mass formula \cite{wghw}.
In the following, we will discuss the problem in
the Noether-charge realization and the Hamiltonian
realization separately.

\subsection{Noether-charge realization}
 The Noether charge defined on the combined Cauchy
surface (\ref{s4.sC}) is combined in the following way:
\begin{eqnarray}\label{s4.sQ1}
Q_\Sigma(\xi) = Q_{\Sigma_1}(\xi)+Q_{\Sigma_2}(\xi),
\end{eqnarray}
where $Q_{\Sigma_1}(\xi)$ and $Q_{\Sigma_2}(\xi)$ are the
Noether charges on $\Si_1$ and $\Si_2$,
respectively.  On shell, $Q_{\Si_1}$ and $Q_{\Si_2}$
may always be expressed, as mentioned before, in terms of
the algebraic summation of the boundary terms
\begin{eqnarray}
\label{s4.sQ2}
Q_{\Sigma_1}(\xi)&=& {\cal Q}_{B_\epsilon}(\xi)-{\cal Q}_{H}(\xi),\\
\label{s4.sQ3}
Q_{\Sigma_2}(\xi) &=&{\cal
Q}_{\infty}(\xi)-{\cal Q}_{B_\epsilon}(\xi),
\end{eqnarray}
where ${\cal Q}_{B_\epsilon}(\xi)$, ${\cal Q}_H(\xi)$, and
${\cal Q}_{\infty}(\xi)$
are evaluated at $B_\epsilon$, the bifurcation
surface $S_H$ and the spatial infinity
$S_\infty$, respectively. It is easy to show that
for the Killing vector field (\ref{s2.kv}),
\be
\label{s4.Q0}
\lim _{\eps \to 0} {\cal Q}_{B_\epsilon}(\chi_K^{\ })
= {\cal Q}_{H}(\chi_K^{\ })
=\frac{\kappa}{8\pi}A.
\ee
That is, ${\displaystyle \lim _{\eps \to 0}}{\cal
Q}_{B_\epsilon}$ for the Killing vector
(\ref{s2.kv}) is proportional to the area $A$ of the
2-dimensional horizon.

According to Sec. III.B, the Noether charges
$Q_{\Sigma_1}(\xi)$ and $Q_{\Sigma_2}(\xi)$ over
$\Si_1$ and $\Si_2$ and their (horizontal)
variation form the algebraic relations for
$\diff$($R \times \Si_1$) and $\diff$($R \times
\Si_2$), respectively.  For example,
the algebraic relation for $Q_{\Sigma_2}(\xi)$
takes the form
\be
\label{s4.Qb1}
\bar \dl_{\xi_2}Q_{\Sigma_2}(\xi_1)-
\bar \dl _{\xi_1}Q_{\Sigma_2}(\xi_2)=
Q_{\Sigma_2}([\xi_1,\xi_2]) + K^Q_{\Si_1}(\xi_2,\xi_2).
\ee
By use of Eq.(\ref{s4.sQ3}), Eq.(\ref{s4.Qb1})
reduces to
\be
\label{s4.Qv2}
\bar \dl _{\xi_2} {\cal Q}_{B_\epsilon}(\xi_1) -
\bar \dl _{\xi_1} {\cal Q}_{B_\epsilon}(\xi_2) -
{\cal Q}_{B_\epsilon}([\xi_1,\xi_2]) -
\tilde K^Q _{B_\epsilon}(\xi_1,\xi_2)  \nonumber \\
= \bar \dl _{\xi_2} {\cal Q}_{\infty}(\xi_1)
 -\bar \dl _{\xi_1} {\cal Q}_{\infty}(\xi_2) -
 {\cal Q}_{\infty}([\xi_1,\xi_2]) -
\tilde K ^Q_{\infty}(\xi_1,\xi_2).
\ee
The two sides of Eq.(\ref{s4.Qv2}) are
evaluated at two different places.  Therefore,
each of them should form an algebra independently.
At most, they are equal to the same constant on
$\Si_2$.  (The value of the constant may be the vector-field
dependent.)  Absorbing the constant into the
central extension $K^Q(\xi_1,\xi_2)$, one has
\be
\label{s4.Qv3}
\bar \dl _{\xi_2} {\cal Q}_{B_\epsilon}(\xi_1)-
\bar \dl _{\xi_1} {\cal Q}_{B_\epsilon}(\xi_2)
={\cal Q}_{B_\epsilon}([\xi_1,\xi_2]) +
K^Q_{B_\epsilon}(\xi_1,\xi_2),
\ee
and
\be
\label{s4.Qv4}
\bar \dl _{\xi_2} {\cal Q}_{\infty}(\xi_1)-
\bar \dl _{\xi_1} {\cal Q}_{\infty}(\xi_2)
 = {\cal
Q}_{\infty}([\xi_1,\xi_2]) + K ^Q_{\infty}(\xi_1,\xi_2).
\ee
Eqs.(\ref{s4.Qv3}) and (\ref{s4.Qv4}) are the algebraic relation
for the partial Noether charges.

\subsection{Hamiltonian realization}
 Since the Hamiltonian functional is also
an additive quantity, it may be written as
\be
H_\Sigma(\xi) = H_{\Sigma_1}(\xi)+H_{\Sigma_2}(\xi).
\ee
On $\Si_i$ ($i=1,2$), the Hamiltonian realization
is
\be
\{ H_{\Si_i}(\xi_1), H_{\Si_i}(\xi_2)\}=H_{\Si_i}([\xi_1, \xi_2])
+ K_{\Si_i}^H(\xi_1,\xi_2).
\ee
If the field ${\bf B}$ exists on the intersection of
$\Sigma_1$ and $\Sigma_2$, the Hamiltonian functionals,
like the Noether charges, may be expressed
in terms of their boundary terms
\be
H_{\Sigma_1}(\xi)&=& J_{B_\epsilon}(\xi)-J_{H}(\xi),\\
H_{\Sigma_2}(\xi) &=&J_{\infty}(\xi)-J_{B_\epsilon}(\xi).
\ee
Similar to the Noether-charge realization,
\be
\{J_H(\xi_1) ,J_H(\xi_2)\}^* &=& J_H([\xi_1, \xi_2]) + K_H^H(\xi_1,\xi_2),\\
\label{s4.SC}
\{J_{B_\eps}(\xi_1) ,J_{B_\eps}(\xi_2)\}^* &=& J_{B_\eps}([\xi_1, \xi_2])
+ K_{B_\eps}^H(\xi_1,\xi_2),\\
\{J_\infty(\xi_1) ,J_\infty(\xi_2)\}^* &=& J_\infty([\xi_1, \xi_2]) +
K_\infty^H(\xi_1,\xi_2).
\ee
Eq.(\ref{s4.SC}) is used to calculate the central term in \cite{c2}.
\omits{For the Hamiltonian realization the similar decomposition
could be made if it might exist.}

\section{Asymptotic Behaviors \label{s5}}

\subsection{Near the spatial infinity}
 In the present paper, we are only interested in the diffeomorphism
of a manifold into itself.  The horizontal variations induced by
such kinds of diffeomorphism do not change the characters of a
manifold.  In particular, a Cauchy surface is mapped into another
Cauchy surface under a horizontal variation in active point of
view.   For stationary asymptotically flat spacetimes, it requires
that the infinite boundaries of Cauchy surfaces coincide at the
same point $i^0$ on the Carter-Penrose diagram.  If the discussion
is confined in the deformation of `$t-r$' plane, it requires that
the vector fields generating diffeomorphisms must satisfy
\be
\xi^a \sim C \left ( \d {\r }{\r t}\right )^a, \qquad \mbox{as $r \to \infty$},
\ee
where $C$ is a constant, $(\frac {\r }{\r t})^a$ is the timelike Killing vector.
The Lie bracket of two vector fields of such a kind obviously
vanishes at the spatial infinity.

\subsection{Near the bifurcation surface $S_H$}

 For the stationary asymptotically flat spacetimes with a black hole, the inner
boundary of the Cauchy surfaces coincide at the bifurcation surface $S_H$.
Limited in the deformation of `$t-r$' plane, it requires that the vector fields
generating diffeomorphisms must satisfy
\be
\xi^a \sim C' \chi_K^a,
\ee
where $C'\sim O(1)$ is another constant, $\chi_K^a$ the Killing vector Eq.(\ref{s2.kv}).
Since $\chi_K^a=0$ on the bifurcation surface, the Lie bracket of two vector
fields of such a kind is also zero at the bifurcation.

\subsection{Near the stretched horizon $B_\eps$}

Let $S=\displaystyle \lim_{\eps \to 0} B_\eps$ be on the Killing horizon.
If $S$ is fixed on the Killing horizon, $\Sigma_1$ and $\Sigma_2$
are the Cauchy surfaces for the spacetime regions $\Sigma_1\times R$
and $\Sigma_2 \times R$, respectively.  In that case, if the discussion is again
limited in `$t-r$' plane,
\be
\xi^a \to 0,  \qquad \mbox{as $\eps \to 0$.}
\ee
Thus, the Lie bracket of the two vector fields is still zero on $S$.

If $S$ is not fixed on the Killing horizon under the diffeomorphism,
$\Sigma_1$ and $\Sigma_2$ are not the Cauchy surfaces of the spacetime
regions $\Sigma_1 \times R$ and $\Sigma_2 \times R$, respectively,
but $\Sigma_1 \bigcup \Si_2$ is still the Cauchy surface of the asymptotically
flat region.  In order $S$ to be mapped from a sphere
to another sphere on the Killing horizon, the vector field should has the
form
\be  \label{s5.vf0}
\xi^a = \tilde T \chi_K^a,  \qquad \mbox{on the Killing horizon}.
\ee
where $\tilde T$ is, at most, the function of $v$, $\th$, and $\phi$,
where $v$ is the parameter of the generator of the Killing horizon and
$\th$ and $\phi$ are coordinates on the cross section of the horizon.
The Lie bracket of two vector
fields of type (\ref{s5.vf0}) gives a new vector field of the same type.
It should be noted that the vector field (\ref{s5.vf0}) does not satisfy
the boundary condition (4.3) in \cite{c2} generally, which reads
\be
\d {\chi_K^a \chi_K^b}{\chi_K^2} \hat \dl_\xi g_{ab} \to 0,  \qquad
\mbox{$\chi_K^2 \to 0$.}
\ee
The boundary condition (4.3) in \cite{c2} is discussed in detail in
Appendix A.

\section{The Null Tetrad and Central Extension \label{s6}}

 Now we are ready to calculate the central extension.  It is obvious that
both the Hamiltonian and Noether-charge realizations are trivial at the spatial
infinity and the bifurcation surface $S_H$.  Even on $S$, both realizations are
also trivial when $S$ is fixed.  In these cases, the central extensions are
all zero.

In this section, we focus on the central extension on $S$ in
the case that $S$ is unfixed under the diffeomorphism.   As was emphasized in
the introduction, in order to estimate the asymptotic behavior precisely we
employ the N-P formalism.

\subsection{Null tetrad}

The (timelike) Killing vector $\chi_K^{\ }$ defines the Killing horizon of a
black hole via $\chi_K^2=0$.  Define a (spacelike) vector field $\rho^a$
orthogonal to the Killing vector $\chi_K^a$ by
\be
\label{s6.Rn}
\del_a \chi_K^2 = -2 \kappa \rho_a,
\ee
where $\kappa$ is the surface gravity on the horizon.  Let $t_i^a$
($i=1,2$) be the two independent tangent vector fields on the cross
section of the horizon.

It is easy to check that $\rho^2 \to 0$, $\rho^a \to \chi_K^a$ and
$\rho^2/\chi_K^2 \to -1$ as the horizon is tended.  These properties
result in that the set of basis consisting of the four vector fields
$\{\chi_K^a, \rho^a, t_i^a\}$ is ill-defined on the horizon.

In order to set up the  basis that are
well-defined on the horizon, we can first choose two real null vector
fields $l^a$, $n^a$ as
\begin{eqnarray}
\label{s6.ln}
l^a&=&\frac{1}{2}(\chi_K^a+\frac{|\chi_K^{\ }|}{\rho}\rho^a)\nonumber\\
n^a&=&-\frac{1}{\chi_K^2}(\chi_K^a-\frac{|\chi_K^{\ }|}{\rho}\rho^a),
\end{eqnarray}
or
\begin{eqnarray}
\chi_K^a&=&l^a-\frac{\chi_K^2}{2}n^a\nonumber\\
\rho^a&=&\frac{\rho}{|\chi_K^{\ }|}(l^a+\frac{\chi_K^2}{2}n^a).
\end{eqnarray}
It is easy to see that $l^an_a=-1$, $l^al_a=n^a n_a=0$
holds on the whole spacetime.  Combined with other two complex null
vector fields $m^a$ and $\bar m ^a$, the set of the null vector fields
$\{l,\ n,\ m,\ \bar m\}$
constitute the well-defined null tetrad fields on the whole
spacetime.

Now, consider the vector fields
\be
\label{s6.vf}
\xi^a=Tl^a+Rn^a =  \tT\chi_K^a+\tR\rho^a .
\ee
Obviously,
\begin{eqnarray}
T&=&\tT+\frac{\rho}{|\chi_K^{\ }|}\tR\nonumber\\
R&=&\frac{\chi_K^2}{2}(-\tT+\frac{\rho}{|\chi_K^{\ }|}\tR).
\end{eqnarray}
If we require both $\tT$ and $\tR$ to be regular functions on
the whole spacetime as required in \cite{c2}, we have
$R \sim O(\chi_K^2)$.

\subsection{The vector-field realization of $\diff$(${\cal M}$) under Lie bracket}

 Let us consider $[l,\ n]$ first.  By definition,
\begin{eqnarray}
\label{s6.[ln]}
[l,\ n]&=& \d 1 2 \left \{[\chi_K^a,\ -\frac{1}{\chi_K^2}\chi_K^a]-[\frac{|\chi_K^{\ }|}{\rho}\rho^a,\
\frac{1}{\chi_K^2}\chi_K^a] +[\chi_K^a,\
\frac{1}{\chi_K^2}\frac{|\chi_K^{\ }|}{\rho}\rho^a] \right .\nonumber \\
&& \left . +[\frac{|\chi_K^{\ }|}{\rho}\rho^a,\
\frac{1}{\chi_K^2}\frac{|\chi_K^{\ }|}{\rho}\rho^a] \right \}
\end{eqnarray}
The first term in the brace bracket in Eq.(\ref{s6.[ln]}) is
\begin{eqnarray}
[\chi_K^a,\ \frac{1}{\chi_K^2}\chi_K^a]&=&\frac{1}{\chi_K^2}[\chi_K^a,\
\chi_K^a]+(D\frac{1}{\chi_K^2})\chi_K^a=0,
\end{eqnarray}
where $D:=\chi_K^a\del_a$.  The second term is
\begin{eqnarray}
[\frac{|\chi_K^{\ }|}{\rho}\rho^a,\
\frac{1}{\chi_K^2}\chi_K^a]
&=&\frac{|\chi_K^{\ }|}{\rho}(\tilde
D\frac{1}{\chi_K^2})\chi_K^a-\frac{1}{\chi_K^2}(D\frac{|\chi_K^{\ }|}{\rho})\rho^a
+\frac{|\chi_K^{\ }|}{\rho}\frac{1}{\chi_K^2}[\rho,\ \chi] \nonumber \\
&=&-2\kappa\frac{1}{\chi_K^2}\frac{\rho}{|\chi_K^{\ }|}\chi_K^a,
\end{eqnarray}
where $\tD:=\rho^a\del_a$.  The third term is
\begin{eqnarray}
[\chi_K^a,\
\frac{1}{\chi_K^2}\frac{|\chi_K^{\ }|}{\rho}\rho^a]=(D\frac{1}{\chi_K^2})\frac{|\chi_K^{\ }|}{\rho}\rho^a
+\frac{1}{\chi_K^2}(D\frac{|\chi_K^{\ }|}{\rho})\rho^a+\frac{1}{\chi_K^2}\frac{|\chi_K^{\ }|}{\rho}[\chi,\ \rho]
=0.
\end{eqnarray}
The last term is
\begin{eqnarray}
[\frac{|\chi_K^{\ }|}{\rho}\rho^a,\
\frac{1}{\chi_K^2}\frac{|\chi_K^{\ }|}{\rho}\rho^a]=
-\frac{\chi_K^2}{\rho^2}(\tilde D\frac{1}{\chi_K^2})\rho^a
+\frac{1}{\chi_K^2}[\frac{|\chi_K^{\ }|}{\rho}\rho^a,\ \frac{|\chi_K^{\ }|}{\rho}\rho^a]
=-2\kappa\frac{1}{\chi_K^2}\rho^a.
\end{eqnarray}
So, the Lie bracket of the null vector fields $l$ and $n$ is
\begin{eqnarray}
[l^a,\ n^a]=-\kappa\frac{\rho}{|\chi_K^{\ }|}n^a\label{[ln]2}.
\end{eqnarray}
For the sake of the later convenience, denote $n^a\del_a$ as $\Dl$ and
$l^a\del_a$ as $\bD$.  Then, the above result is equivalent to
\begin{eqnarray}
[\bD,\ \Dl]=-\kappa\frac{\rho}{|\chi_K^{\ }|}\Dl
\end{eqnarray}
for any function $f$.
In comparison with the commutator of the N-P formalism \cite{exa}, the
following relation among the N-P coefficients are obtained:
\begin{eqnarray}
\label{s6.npc}
\tau+\bar\pi &=&0, \nonumber\\
\gamma+\bar\gamma &=&0, \nonumber\\
\eps+\bar\eps &=&\kappa\frac{\rho}{|\chi_K^{\ }|}.
\end{eqnarray}
For the vector fields of Eq.(\ref{s6.vf}),
\begin{eqnarray}
\label{s6.x1x2}
[\xi^a_1,\ \xi^a_2]&=&[T_1l^a,\ T_2l^a]+[R_1n^a,\ T_2l^a]+[T_1l^a,\ R_2n^a]+[R_1n^a,\ R_2n^a]\nonumber\\
&=&(T_1\bD T_2-T_2\bD T_1)l^a+(R_1\Dl T_2-R_2\Dl T_1)l^a+(T_1\bD R_2-T_2\bD
R_1)n^a\nonumber\\
&{ }&+\kappa\frac{\rho}{|\chi_K^{\ }|}(R_1T_2-T_1R_2)n^a+(R_1\Dl
R_2-R_2\Dl R_1)n^a.
\end{eqnarray}
Because
\begin{eqnarray}
\label{s6.dchi}
\bD\chi_K^2&=&\frac{1}{2}(\chi_K^a+\frac{|\chi_K^{\ }|}{\rho}\rho^a)\del_a\chi_K^2
=\chi_K^2\kappa\frac{\rho}{|\chi_K^{\ }|}=O(\chi_K^2)\nonumber\\
\Dl\chi_K^2&=&-\frac{1}{\chi_K^2}(\chi_K^a-\frac{|\chi_K^{\ }|}{\rho}\rho^a)\del_a\chi_K^2
=2\kappa\frac{\rho}{|\chi_K^{\ }|}=O(1),
\end{eqnarray}
it is easy to see that the Lie bracket of $\xi$'s is closed as long as $R =O(\chi_K^2)$.

\subsection{On the central extension}

When $S$ is mapped from one sphere to another on the Killing
horizon under the diffeomorphism, either $\Si_1$ or $\Si_2$ cannot
be regarded as the Cauchy surface of subsystems $\Si_1\times R$
and $\Si_2 \times R$ and the Cauchy surface of the whole
asymptotically flat region as well.  The legitimacy of applying
Eq.(\ref{s3.dH0}) to the subsystem becomes questionable because
the integral over the (partial) Cauchy surface is taken in
Eq.(\ref{s3.dH0}). \omits{As mentioned in Sec. III.A, the
condition of Eq.(\ref{s3.dH}) is $\hat \dl_{\xi_2}\xi_1=0$.}  In
this subsection,we do not plan to discuss the problems in the
formulation.  Instead, we will just estimate the central extension
in the null tetrad according to Ref. \cite{c2}.

It has been written in \cite{c2} that for vacuum GR 
\be
\label{s6.hb0}
\{J_{B_\eps}(\xi_1),J_{B_\eps}(\xi_2)\}^* =
\d 1 {16 \pi} \int_{B_\eps} {\mbox {\boldmath $\eps$}}_{abcd}
[\xi^c_2\del_e(\del^e\xi^d_1-\del^d\xi^e_1)
- \xi^c_1\del_e(\del^e\xi^d_2-\del^d\xi^e_2)].
\ee
For a class of vector fields considered in \cite{c2},
\be \label{s6.xib}
[\xi_1, \xi_2] \cdot {\bf B} =0
\ee
on the boundary.  Thus, the boundary term of the Hamiltonian functional reads
\be \label{hlb}
J_{B_\eps}([\xi_1,\xi_2])={\cal Q}_{B_\eps}([\xi_1,\xi_2])= -\frac{1}{16 \pi}\int _{B_\eps}
{\mbox {\boldmath $\epsilon$}} _{abcd}\del^c
(\xi_1^e \del_e \xi_2^d - \xi_2^e \del_e \xi_1^d ),
\ee
and then the possible central term is
\be
\label{kh}
K_S^H(\xi_1,\xi_2) &=& \lim_{\eps \to 0}\left (\{J_{B_\eps}(\xi_1),J_{B_\eps}(\xi_2)\}^*
- J_{B_\eps}([\xi_1,\xi_2]) \right )\nonumber \\
&=& \lim_{\eps \to 0} \frac{1}{8 \pi}\int _{B_\eps}
{\mbox {\boldmath $\epsilon$}} _{abcd} \left [
\del _e (\xi_{[1}^c \del ^d \xi_{2]}^e) -
2 \xi_{[1}^{[c} \del ^{e]} \del_{|e|} \xi_{2]}^{d}
\right ].
\ee

For the vector field (\ref{s6.vf}), the right-hand side of Eq.(\ref{s6.hb0}) becomes
\begin{eqnarray}
\label{s6.q12ln}
\d 1 {16\pi}\int_{B_\eps}{\mbox {\boldmath $\epsilon$}}_{abcd}
\{(T_2l^c+R_2n^c)\del_e[\del^e(T_1l^d+R_1n^d)-\del^d(T_1l^e+R_1n^e)] \\
-[1\leftrightarrow 2 ]\}
\end{eqnarray}
The first term of the integrand in Eq.(\ref{s6.q12ln}) is
\begin{eqnarray}
&& {\mbox {\boldmath $\epsilon$}}_{abcd}\{(T_2l^c+R_2n^c)\del_e\del^e(T_1l^d+R_1n^d)
-[1\leftrightarrow 2 ]\}\nonumber\\
&=&{\mbox {\boldmath $\epsilon$}}_{abcd} \{2T_2l^c(\del_eT_1)\del^el^d+
{2R_2n^c(\del_eR_1)\del^en^d}+T_2l^c(\del_e\del^eR_1)n^d\nonumber\\
&&\qquad +2T_2l^c(\del_eR_1)\del^en^d+{R_1l^cT_2\del_e\del^en^d}+
{R_2n^c(\del_e\del^eT_1)l^d}\nonumber\\
&&\qquad +{2R_2n^c(\del_eT_1)\del^el^d}+{R_2T_1n^c\del_e\del^el^d}
-[1\leftrightarrow 2 ]\}\nonumber\\
&=&{\mbox {\boldmath $\epsilon$}}_{abcd} \{2T_2l^c(\del_eT_1)\del^el^d+
T_2l^c(\del_e\del^eR_1)n^d+2T_2l^c(\del_eR_1)\del^en^d \nonumber \\
&& \qquad -[1\leftrightarrow 2 ] \} + O(\chi_K^2).
\end{eqnarray}
Similarly, the second term is
\begin{eqnarray}
&&{\mbox {\boldmath $\epsilon$}}_{abcd}\{(T_2l^c+R_2n^c)\del_e\del^d(T_1l^e+R_1n^e)
-[1\leftrightarrow 2 ]\}\nonumber\\
&=&{\mbox {\boldmath $\epsilon$}}_{abcd}\{T_2l^c(\del_e\del^dT_1)l^e+T_2l^c(\del^dT_1)\del_el^e
+T_2l^c(\del_eT_1)\del^dl^e+T_2l^c(\del_e\del^dR_1)n^e\nonumber\\
&&\qquad +T_2l^c(\del^dR_1)\del_en^e+T_2l^c(\del_eR_1)\del^dn^e
-[1\leftrightarrow 2 ]\} + O(\chi_K^2).
\end{eqnarray}
Now, Eq.(\ref{s6.q12ln}) becomes
\begin{eqnarray} \label{s6.I}
\omits{&&(\hat \dl _{\xi_2} {\cal Q}_{B_\epsilon})(\xi_1)-
(\hat \dl _{\xi_1} {\cal Q}_{B_\epsilon})(\xi_2) \nonumber \\}
\mbox{Eq.(\ref{s6.q12ln})}&=&\d 1 {16\pi}\int_{B_\eps}{\mbox {\boldmath $\epsilon$}}_{abcd}
\{[2T_2l^c(\del_eT_1)\del^el^d+T_2l^c(\del_e\del^eR_1)n^d+2T_2l^c(\del_eR_1)\del^en^d\nonumber\\
&{ }&\qquad \qquad -T_2l^c(\del_e\del^dT_1)l^e-T_2l^c(\del^dT_1)\del_el^e
-T_2l^c(\del_eT_1)\del^dl^e\nonumber\\
&{ }&\qquad \qquad -T_2l^c(\del_e\del^dR_1)n^e-T_2l^c(\del^dR_1)\del_en^e
-T_2l^c(\del_eR_1)\del^dn^e]\nonumber\\
&{ }&\qquad \qquad -[1\leftrightarrow 2 ]\} + O(\chi_K^2).
\end{eqnarray}

In the N-P formalism, we have $g_{ab}=-l_an_b-n_al_b+m_a\bar m_b+\bar m_am_b$,
$\dl ^a_b=-l^an_b-n^al_b+m^a\bar m_b+\bar m^am_b$, and
\be
\del_a = -l^a \Dl -n^a \bD + \bar m^a \dl + m^a \bar \dl,
\ee
where $\dl =m^a \del_a$ and $\bar \dl = \bar m^a \del_a$.  These lead to
\begin{eqnarray}
\del_a\del^a&=&-\bD\Dl -\Dl \bD+\bar\dl\dl+\dl\bar\dl
-[(\eps+\bar\eps)-(\varrho+\bar\varrho)]\Dl \nonumber\\
&{ }&+[(\gamma+\bar\gamma)-(\mu+\bar\mu)]\bD+
(\pi-\bar\tau-\alpha+\bar\beta)\dl+(\bar\pi-\tau-\bar\alpha+\beta)\bar\dl \nonumber \\
&=&-\bD\Dl -\Dl \bD+\bar\dl\dl+\dl\bar\dl-[\kappa\frac{\rho}{|\chi_K^{\ }|}-(\varrho+\bar\varrho)]\Dl
-(\mu+\bar\mu)\bD\nonumber\\
&{}&+(\pi-\bar\tau -\alpha +\bar\beta)\dl+(\bar\pi -\tau-\bar\alpha+\beta)\bar\dl,
\end{eqnarray}
where Eq.(\ref{s6.npc}) has been used in the second equality.
In addition, ${\mbox {\boldmath $\epsilon$}}_{abcd}=i{\bf l}_a\wedge {\bf n}_b\wedge
{\bf m}_c\wedge{\bf \bar m}_d$.

Now, we may much more precisely analyze Eq.(\ref{s6.I}) term by term
under the assumptions that
\be \label{s6.cond1}
R, \ \dl R, \ \bar \dl R \sim O(\chi_K^2)
\ee
and
\be \label{s6.cond2}
\varrho + \bar \varrho = -  \bar m^a m^b \del_a l_b - m^a \bar m^b \del_a l_b
\sim O(\chi_K^2).
\ee
Since the volume element on
$B_{\eps}$ is ${\mbox {\boldmath $\epsilon$}}_{ab}=i{\bf m}_a\wedge{\bf \bar m}_b$,
only those terms in the brace bracket in Eq.(\ref{s6.I}), which contains ${\bf l}
\wedge {\bf n}$,
will contribute to the integral.  So, we will only write out
those terms in following calculation and denote those equalities
as `$\eqo$'.
In the calculation, the relations for the null tetrad in Appendix B will be used.
The first term is
\begin{eqnarray}
2T_2l^c(\del_eT_1)\del^el^d&=& 2T_2l^c(-l^dn_f-n^dl_f+m^d\bar m_f+\bar
m^dm_f)(\del_eT_1)\del^el^f\nonumber\\
&\eqo&-2T_2l^cn^dl_f(\del_eT_1)\del^el^f\nonumber\\
&=&-l^cn^dT_2(\del_eT_1)\del^el^2=0.
\end{eqnarray}
The second term is
\begin{eqnarray}
T_2l^cn^d\del_e\del^eR_1
&=&l^cn^dT_2\{-\bD\Dl -\Dl \bD-
\kappa\frac{\rho}{|\chi_K^{\ }|}\Dl +(\varrho+\bar \varrho)\Dl \}R_1 .
\end{eqnarray}
The third term is
\begin{eqnarray}
2T_2l^c(\del_eR_1)\del^en^d
&\eqo&-2T_2l^cn^d(\del_eR_1)l_f\del^en^f\nonumber\\
&=&l^cn^d(2\kappa\frac{\rho}{|\chi_K^{\ }|}T_2\Dl R_1) + O(\chi_K^2).
\end{eqnarray}
The forth term is \cite{note}
\begin{eqnarray}
-T_2l^c(\del_e\del^dT_1)l^e
&\eqo& T_2l^cn^dl^el_f\del_e \del^fT_1\nonumber\\
&=&l^cn^d[T_2\bD^2T_1+ T_2 (l^en^f\del_e l_f)\bD T_1
-T_2 (l^e \bar m^f\del_e l_f)\dl T_1 \nonumber \\
&&\qquad \qquad  - T_2 (l^e m^f\del_e l_f)\bar \dl T_1]
\nonumber \\
&=&l^cn^d[T_2\bD^2T_1-
\ka \d \rho {|\chi_K^{\ }|} T_2 \bD T_1  +\bar \iota T_2 \dl T_1 + \iota T_2 \bar\dl T_1].
\end{eqnarray}
The fifth term is
\begin{eqnarray}
-T_2l^c(\del^dT_1)\del_el^e
&\eqo& l^cn^d T_2(\bD T_1)\del_el^e\nonumber\\
&=& l^cn^d(\kappa\frac{\rho}{|\chi_K^{\ }|}T_2\bD T_1)+O(\chi_K^2),
\end{eqnarray}
The sixth term is
\begin{eqnarray}
-T_2l^c(\del_eT_1)\del^dl^e
&\eqo&l^cn^dT_2(\del_eT_1)l_f\del^fl^e\nonumber\\
&=& l^cn^dT_2 (\kappa \d \rho {|\chi_K^{\ }|}T_2\bD T_1
 -\iota \bar \dl T_1 -\bar \iota \dl T_1).
\end{eqnarray}
The seventh term is
\begin{eqnarray}
-T_2l^c(\del_e\del^dR_1)n^e
&\eqo& l^cn^dT_2n^el_f\del_e\del^fR_1\nonumber\\
&=&l^cn^d(T_2\Dl \bD R_1)+O(\chi_K^2).
\end{eqnarray}
The eighth term is
\begin{eqnarray}
-T_2l^c(\del^dR_1)\del_en^e
&\eqo& l^cn^dT_2(\bD R_1)\del_en^e = O(\chi_K^2).
\end{eqnarray}
The last term is
\begin{eqnarray}
- T_2l^c(\del_eR_1)\del^dn^e
&\eqo&l^cn^d T_2 (\del_eR_1)l_f\del^fn^e\nonumber\\
&=& - l^cn^d (\ka \d \rho {|\chi_K^{\ }|} T_2\Dl R_1)
+ O(\chi_K^2).
\end{eqnarray}
Thus, 
\begin{eqnarray}
\omits{(\hat \dl _{\xi_2}{\cal Q})(\xi_1)-(\hat \dl _{\xi_1}{\cal Q}(\xi_2)}
\mbox{Eq.(\ref{s6.q12ln})}&=&\d 1 {16\pi}\int_{B_\eps}{\mbox {\boldmath $\epsilon$}}_{abcd}l^cn^d
\{[T_2(-\bD\Dl -\Dl \bD-\kappa\frac{\rho}{|\chi_K^{\ }|}\Dl )R_1  \nonumber\\
&{ }& \qquad +2\kappa\frac{\rho}{|\chi_K^{\ }|}T_2\Dl R_1+T_2\bD^2T_1
-\kappa \frac{\rho}{|\chi_K^{\ }|} T_2\bD T_1  \nonumber\\
&{ }& \qquad + \kappa\frac{\rho}{|\chi_K^{\ }|}T_2\bD T_1
+\kappa \frac{\rho}{|\chi_K^{\ }|} T_2\bD T_1+T_2\Dl \bD R_1-
\kappa\frac{\rho}{|\chi_K^{\ }|}T_2\Dl R_1]\nonumber\\
&{ }& \qquad  + O(\chi_K^2)%
-[1\leftrightarrow 2 ]\}\nonumber\\
&=&\d 1 {16\pi}\int_{B_\eps}{\mbox {\boldmath $\epsilon$}}_{abcd}l^cn^d
\{[-T_2\bD\Dl R_1+T_2\bD^2T_1+\kappa \frac{\rho}{|\chi_K^{\ }|} T_2\bD
T_1] \nonumber \\
&& \qquad +O(\chi_K^2)-[1\leftrightarrow 2]\}.
\end{eqnarray}

On the other hand, the boundary term of the Hamiltonian functional is
equal to the partial Noether charge for the given vector fields, which may be
written on shell as
\begin{eqnarray}
J_{B_\eps}(\xi)={\cal Q}_{B_\eps}(\xi)&=&-\d 1 {16\pi}\int_{B_\eps}*d{\mbox {\boldmath $\xi$}}\nonumber\\
&=&-\d 1 {16\pi}\int_{B_\eps}*(dT \wedge {\bf l} + T d{\bf l}
+d R \wedge {\bf n} + R d{\bf n})\nonumber\\
&=&-\d 1 {16\pi}\int_{B_\eps}*(dT \wedge {\bf l} + T d{\bf l}
+d R \wedge {\bf n}) + O(\chi_K^2).
\end{eqnarray}
Since
\begin{eqnarray}
dT\wedge{\bf l}&\eqo&(\bD T){\bf l}\wedge{\bf n}\\
(d{\bf l})_{ab} 
&\eqo& -2n_{[a} \bD l_{b]} -2l_{[a}\Dl l_{b]} \nonumber\\
&\eqo& - (n^c \bD l_{c}) {\bf l}\wedge{\bf n} \nonumber \\
&=&\kappa \d \rho {|\chi_K^{\ }|}{\bf l}\wedge{\bf n}+O(\chi_K^2)\\
dR\wedge{\bf n}&\eqo&-(\Dl R){\bf l}\wedge{\bf n},
\end{eqnarray}
the boundary term of the Hamiltonian functional with respect to the vector
field takes the form of
\begin{eqnarray}
\label{qxi}
J_{B_\eps}(\xi)&=&-\d 1 {16\pi}\int_{B_\eps}{\mbox {\boldmath $\epsilon$}}_{abcd}l^cn^d
(\bD T+\kappa \d \rho {|\chi_K^{\ }|}T-\Dl R) + O(\chi_K^2).
\end{eqnarray}
Thus,
\begin{eqnarray} \label{s6.jx1x2}
J_{B_\eps}([\xi_1,\ \xi_2]) &=& -\d 1 {16\pi}\int_{B_\eps}{\mbox {\boldmath $\epsilon$}}_{abcd}
l^cn^d[\bD(T_1\bD T_2+R_1\Dl T_2) +\kappa\d \rho {|\chi_K^{\ }|}(T_1\bD T_2+R_1\Dl T_2)\nonumber\\
&{ }&\qquad \qquad -\Dl (T_1\bD R_2+\kappa\frac{\rho}{|\chi_K^{\ }|}R_1T_2+R_1\Dl R_2)]-
(1\leftrightarrow 2 )\nonumber\\
&=& -\d 1 {16\pi}\int_{B_\eps}{\mbox {\boldmath $\epsilon$}}_{abcd}
l^cn^d (T_1\bD^2 T_2+\kappa \frac{\rho}{|\chi_K^{\ }|}T_1\bD T_2-T_1\bD\Dl R_2)+O(\chi_K^2) \nonumber \\
&&\qquad \qquad -(1\leftrightarrow 2).
\end{eqnarray}
Therefore,
\begin{eqnarray}
\{J_S(\xi_1), J_S({\xi_2}) \} ^* - J_S([\xi_1,\ \xi_2])=0.
\end{eqnarray}
This means that even when $S$ is mapped from one point to another on the Killing horizon
as considered in \cite{c2},  {\it the central extension is also zero} as long as
the conditions (\ref{s6.cond1}) and (\ref{s6.cond2}) are satisfied!

It should be mentioned that in \cite{c2} the
asymptotic conditions had been proposed to
specify a subset of the class of vector fields
for getting the non-vanishing center extension.
However, those vector fields do satisfy
conditions (\ref{s6.cond1}) and (\ref{s6.cond2}). This
will be shown in details in Appendix A.

Since the boundary terms of the Hamiltonian functionals
coincide with the partial Noether charges, all the
calculations in this subsection are available to the
Noether-charge realization.  Namely, the central
extension in the Noether-charge realization also
vanishes \cite{ghw}.

\section{Comment on Virasoro Algebra \label{s7}}

Now, we turn to discuss the realization of the subalgebra
of the $\diff(\cal M)$ algebra on the boundary $S$
in the basis of $\{\chi_K^a,\ \rho^a,\ t_1^a,\ t_2^a\}$.

\subsection{The vector fields and the boundary condition}

Following Ref. \cite{c2}, consider the vector fields
\be
\label{s7.xi}
\xi^a = \tilde T \chi_K^a + \tilde R\rho^a
\ee
with
\be
\label{s7.R}
\tilde R = \d 1 \kappa \d {\chi_K^2}{\rho^2}
\chi_K^a \del _a \tilde T
\ee
and
\be
\label{s7.T'}
\rho^a\del_a \tilde T |_H= 0.
\ee

Since the tangent vector fields on a spacetime manifold with either
Lorentzian signature or Euclidean signature are all {\it real} vector fields,
$\tilde T$ must be a real function.  In order to obtain the subalgebra on
the boundary $S$, $\tilde T$ may be further chosen as
\begin{eqnarray} \label{s7.An}
A_n=\d 1 \ka \cos[n(\ka v+\varphi)]
\end{eqnarray}
and
\begin{eqnarray} \label{s7.Bn}
B_n=\d 1 \ka \sin[n(\ka v+\varphi)],
\end{eqnarray}
where $v$ is the parameters of the integral curves of $\chi_K^a$
such that $\chi_K^a \del_a v=1$, the integers $n \geq 0 $, 
and $\varphi$ is the coordinate on $B_\eps$ with $2\pi$ period.
Obviously, the complex combinations of $A_n$ and $B_n$,
\begin{eqnarray}
\tilde T_n=A_n \pm iB_n=\d 1 \ka e^{i n(\ka v +\varphi )},
\end{eqnarray}
has the form of
\begin{eqnarray} \label{s7.Tn}
\tilde T_n=\d 1 \ka e^{i n \ka v} f_n , \qquad \mbox {with $n \in Z$,}
\end{eqnarray}
where $f_n, ~\forall n \in Z$, satisfy
\begin{eqnarray} \label{s7.onc}
\int _{S}{\mbox {\boldmath $\hat \epsilon$}}_{ab}f_mf_n =
\dl_{m+n,0}.
\end{eqnarray}
Eq.(\ref{s7.Tn}) with Eq.(\ref{s7.onc}) has been used to obtain the Virasoro
algebra in \cite{c2}.

\subsection{Realization of Virasoro algebra by vector fields}

In \cite{c2}, the Lie bracket of two vector fields of type (\ref{s7.xi})
was
\begin{eqnarray} \label{s7.Lie1}
\ [ \xi_1,\xi_2]^a &=& (\tilde T_1 D \tilde T_2 - \tilde T_2 D \tilde T_1)\chi_K^a 
+\d 1 \kappa \d {\chi_K^2}{\rho^2} D(\tilde T_1 D \tilde T_2 -\tilde T_2 D \tilde T_1)
\rho^a.
\end{eqnarray}
When $\tilde T = A_m$ and $B_m$, Eq.(\ref{s7.Lie1}) is given by
equivalent to
\begin{eqnarray} \label{s7.Var1}
\ [ A_m, A_n ]&=&\kappa(mB_mA_n-nA_mB_n),\\
\ [ B_m, B_n ]&=&-\kappa(mA_mB_n-nB_mA_n),\\
\ [ A_m, B_n ]&=&\kappa(nA_mA_n+mB_mB_n),
\end{eqnarray}
which lead to
\begin{eqnarray} \label{s7.Var2a}
\ [ A_m, A_n ]-[B_m,B_n]=(m-n) B_{m+n}, \\
\label{s7.Var2b}
\ [ A_m, B_n ]-[A_n,B_m]=(m-n) A_{m+n}.
\end{eqnarray}
They are equivalent to
\begin{eqnarray}\label{s7.Va1}
i[ \tilde T_m, \tilde T_n ] = (m - n) \tilde T_{m+n}.
\end{eqnarray}
Eq.(\ref{s7.Va1}) is nothing else but the complex expression of the classical
Virasoro algebra \cite{cf}, (more precisely, the Witt algebra $\diff({\cal S}^1)$
\cite{wa},)  while (\ref{s7.Var2a}) and (\ref{s7.Var2b}) are the
real expressions for the same algebra.  Therefore, $\tilde T_n$
(or equivalently, the set of $A_n$ and $B_n$) chosen in such a
way would form a representation of the classical Virasoro algebra if the
condition (\ref{s7.T'}) was reasonable and if Eq.(\ref{s7.Lie1}) was valid.

It has been argued in \cite{dgw} that the condition (\ref{s7.T'}) implies
$\tilde D\tilde T=0$ on the horizon which is conflict to the basic
requirement $D\tilde T \neq 0$.  
According to the discussion in the Sec. VI and
Appendix A, the Lie bracket of the two vector fields of type (\ref{s7.xi})
should be
\begin{eqnarray}
[\xi^a_1,\ \xi^a_2]=(T_1\bD T_2-T_2\bD T_1)l^a+(R_1\Dl T_2-R_2\Dl T_1)l^a +O(\chi_K^2),
\end{eqnarray}
which leads to
\begin{eqnarray} \label{s7.Lie2}
[\xi^a_1,\ \xi^a_2]&=& \d 1 2[(\tilde T_1 + \d \rho {|\chi|} \tilde R_1)
(D+\d {|\chi|} \rho \tD )(\tilde T_2 + \d \rho {|\chi|} \tilde R_2) ]l^a \nonumber \\
&& + \d 1 2 \chi^2[(-\tilde T_1 + \d \rho {|\chi|} \tilde R_1)
\d 1 {\chi^2}(-D+\d {|\chi|} \rho \tD )(\tilde T_2 + \d \rho {|\chi|} \tilde R_2)] l^a
- (1 \leftrightarrow 2) +O(\chi_K^2) \nonumber \\
&=&\d 1 2 (\tilde T_1 D \tilde T_2- \tilde T_2 D \tilde T_1) \chi^a +
\d 1 2 \d 1 \kappa \d {\chi^2}{\rho^2}D(\tilde T_1 D \tilde T_2- \tilde T_2 D \tilde T_1)
\rho^a \nonumber \\
&& -\d 1 2 \d 1 \kappa \d {|\chi^2|} \rho D (\tilde T_1 D \tilde T_2- \tilde T_2 D \tilde T_1) \chi^a
+\d 1 2 \d  {|\chi^2|} \rho (\tilde T_1 D \tilde T_2- \tilde T_2 D \tilde T_1) \rho^a
+O(\chi_K^2)
\end{eqnarray}
under the condition (\ref{s7.T'}).  It is obviously different from Eq.(\ref{s7.Lie1}).
Hence, even for the choice (\ref{s7.Tn}), Eq.(\ref{s7.Lie2}) does not
give the Virasoro algebra.

\subsection{The Hamiltonian realization of Virasoro algebra}

\subsubsection{The non-zero Hamiltonian functionals}

 By definition,
\be \label{s7.n0Q}
{\cal Q}_S(\tilde T_n) = \frac 1 {16\pi}\int_S {\mbox {\boldmath $\hat \eps$}}_{ab}
\left (2\kappa \tilde T_n - \frac 1 \kappa D^2 \tilde T_n \right )
=  \frac {2+n^2} {16\pi} A_H\dl_{n,0} = \frac {A_H} {8\pi} \dl_{n,0}.
\ee
Namely, the partial Noether charges and thus the boundary terms of the
Hamiltonian functionals with respect to the vector
fields chosen in the above way all vanish except ${\cal Q}_S(\tilde T_0)$
and $J_S(\tilde T_0)$, which are equal to $A_H/8\pi$.

In the Virasoro algebra, $\{L_n\}$ are the generators of the conformal
symmetry, which
should not be equal to zero.  In the present case, the boundary terms of the
Hamiltonian functionals act as the generators of the conformal symmetry.
The vanishing 
boundary terms of the Hamiltonian functionals imply the Virasoro algebra
is trivial.  Namely, when $n, m \neq 0$ and $n \neq -m$,
\be
\{J_S(T_m), J_S(T_n)\} ^*
= (m-n)J_S(\tilde T_{m+n})
\ee
reads $0=0$ actually.

\subsubsection{The surface densities 
the classical Virasoro algebra}

 The problem may not be so serious because the integrands of the
boundary terms of the Hamiltonian functionals
may form the required algebraic relation.  Unfortunately,
the following calculation shows that the surface densities
with respect to the vector fields chosen above do not
agree with the 
classical Virasoro algebra.

For the vector fields (\ref{s7.xi}) with (\ref{s7.R}) and (\ref{s7.T'}), Eq.(\ref{s6.hb0})
in the (ill-defined) basis $\{\chi_K^a,\ \rho^a,$ $t_1^a,\ t_2^a\}$ gives rise
to \cite{note2}
\begin{eqnarray}
\label{s7.Qb5}
\{J_S(\tilde T_m), J_S(\tilde T_n)\} ^*
& = &\d 1 {16\pi} \int_S {\mbox {\boldmath $\hat \eps$}}_{ab}
[\d 1 \kappa (D\tilde T_n D^2 \tilde T_m - D\tilde T_m D^2 \tilde T_n) \nonumber \\
&& \qquad \qquad - 2\kappa(\tilde T_n D \tilde T_m- \tilde T_m D\tilde T_n)]
\nonumber \\
& = &-i(m-n)J_S(\tilde T_{m+n}) +i \d {(m^3-n^3)\ka} {16\pi}
\int_S {\mbox {\boldmath $\hat \eps$}}_{ab} \tilde T_{m+n}
\end{eqnarray}
It looks like the classical Virasoro algebra\omits{\be  \label{QVar1}
&& \left [ (\hat \dl _{A_n} {\cal Q}_S)(A_m) - (\hat \dl _{A_m}
{\cal Q}_S)(A_n) \right ] -\left [ (\hat \dl _{B_n} {\cal Q}_S)(B_m)
 - (\hat \dl _{B_m} {\cal Q}_S)(B_n)\right ] \nonumber\\
& =& (m-n) {\cal Q}_S(B_{m+n})
\ee (See: Appendix C)}
because $\int_S{\mbox {\boldmath $\hat \eps$}}_{ab} \tilde T_{m+n}=0$.
Unfortunately, $J_S(\tilde T_{m+n})=0$ (as well as
$J_S(\tilde T_{m})=0, \forall \ m\neq 0$
) on the same footing as $\int_S{\mbox {\boldmath $\hat \eps$}}_{ab} \tilde T_{m+n}=0$.
Thus, Eq.(\ref{s7.Qb5}) cannot be regarded as the Virasoro algebra.
It is obvious from Eq.(\ref{s7.Qb5}) that the surface densities of the boundary terms
of the Hamiltonian functionals do not fulfill the classical Virasoro
algebra.
Therefore, it is difficult to convince that Eq.(\ref{s7.Qb5}) 
is the realization of a Virasoro algebra because all terms deviating from the
Virasoro algebra as well as all boundary terms of Hamiltonian functionals
except one vanish on the same footing.

\section{Conclusion and Discussion}

 In a diffeomorphism invariant theory of gravity, the Noether
charges themselves, as in other classical and quantum field
theories, may be used to realize the symmetry (in the present
case, the diffeomorphism invariance) of the theory.  This is a
complete covariant approach without using the Poisson bracket,
which is well defined in the canonical formalism but still
contains some ambiguities in the covariant formalism, at least,
for the horizontal variations. It is emphasized that the Noether
charge on shell are well defined for any given boundary condition
and any given vector field and may always be expressed on shell in
terms of the algebraic summation of the boundary terms on a Cauchy
surface \cite{{iw},{wghw}}.  The latter property results in the
Noether-charge realization being separated into the realization of
the partial Noether charge of the two-dimensional closed surface
that is the boundary of the Cauchy surface.

For the Killing vectors, the Noether charge approach may give rise
to certain relations among the Noether charges with respect to the
symmetries generated by the Killing vectors. For the stationary
axisymmetric spacetimes with a black hole, the vacuum black hole
mass formula as a whole can be viewed as a (total) Noether charge
for the combination of the Noether charges for the Killing vectors
in GR. The first law of black hole mechanics can also be re-derived
within the Noether charge formalism under certain conditions.  In
addition, only the horizontal variations are needed in the Noether
charge approach. This is also in agreement with the spirit of
Noether's theorem.

In vacuum GR and when $\xi \cdot {\bf B}=0$ on the boundary, the
Noether-charge realization should coincide with the Hamiltonian
realization because in this case the boundary terms of the
Hamiltonian functional are equal to the partial Noether charges of
boundary surface. Unfortunately, Hamiltonian functionals do not
always exist because they are related to Noether charges by Eq.
(\ref{s3.HQr}) with the definition of ${\bf B}$ given in
variational equation (\ref{s3.B}) and because the variational
equation does not always have a solution for a given boundary
condition and a vector field. This problem, in fact, has been
pointed out by Ward and his collaborator \cite{{wald},{iw}}.
Obviously, if the 3-form ${\bf B}$ does not exist, one cannot
define the Hamiltonian functional and thus the Hamiltonian
realization of $\diff(\cal M$) algebra.

As mentioned before, the Hamiltonian realization has not well set
up because the problem in the definition of the Poisson bracket in
the covariant phase-space formalism has not yet been solved
without debate. Discarding this problem, we may conclude from
Eqs.(\ref{s4.SC}), (\ref{s6.hb0}) and (\ref{s6.jx1x2}) that the
central extension on the Killing horizon for the large class of
vector fields including those studied in \cite{c2} should vanish,
even though in that case the boundary of $\Si_1$ or $\Si_2$ is not
fixed in the variation so that they cannot be treated as the
Cauchy surfaces of portions of the manifold.  The main reason that
our conclusion is different from
\cite{{c2},{c3}},\cite{jy1}-\cite{c4} is that the null tetrad is
used here, which is well-defined on the horizon, while in the
previous works the vector fields $\{\chi_K^a,\ \rho^a,\ t_1^a,\
t_2^a\}$ are treated as a set of basis, which are ill-defined on
the horizon. The vanishing central term implies that the black
hole entropy cannot be explained by such a kind of the classical
symmetry analysis on the horizon.

It should be mentioned, however, that in the abstract and the
section I all the phrases ``central term " are always with the
quotation mark. This is due to the fact that it is not really the
central term in usual sense\omits{ of the projective
representation of Lie algebra}. It is in fact the one proportional
to the partial Noether-like charge of the Noether-like current in
(\ref{s3.ccs}). In order to justify whether the one corresponding
to the entire Norther-like charge is vanishing, it is needed to
consider the whole Cauchy surface $\Sigma$ from the bifurcation
surface $S_H$ to the spacial infinity $i^0$ rather than the
partial Cauchy surface. But, even for the null-frame has been used
in this paper there are still some flaws to be solved. In
addition, the event horizon is not a real boundary of the
spacetime with a black hole. Therefore, it is also questionable to
merely consider the Hamiltonian functionals \omits{partial Noether
charge }with respect to the partial Cauchy surface $\Sigma_2$. How
to solve these problems are still under investigation.

Although the algebra of the vector fields satisfying
(\ref{s7.xi}), (\ref{s7.R}) and (\ref{s7.T'}) under the Lie
bracket seems to be equivalent to the classical Virasoro algebra
in the ill-defined basis $\{\chi_K^a,\ \rho^a,\ t_1^a,\ t_2^a\}$,
the analysis in the N-P formalism shows that it does not give a
Virasoro algebra.  Even in the sense of \cite{c2} that the Lie
bracket algebra of the vector fields is equivalent to the Virasoro
algebra on the horizon, the algebra of the corresponding boundary
terms of Hamiltonian functionals is trivial because all the
boundary terms except one vanish.  The surface densities of the
boundary terms do not form the algebraic relation of the Virasoro
algebra.  It again challenges the proposal that the origin of the
black hole entropy can be explained by the classical symmetry
analysis on the horizon.

Finally, as was mentioned at beginning, though the present discussion is
confined in 4 dimension, it is easy to generalize the discussion to any
dimension in principle.

\section*{Acknowledgments}

We would like to thank Professors/Drs.
C.-B. Liang,  Y. Ling, R.-S. Tung,  K. Wu, M. Yu
Z. Chang, B. Zhou, and C.-J. Zhu for helpful discussion.
This project is in part supported by National Science
Foundation of China under Grants Nos. 90103004,
10175070.  One of the authors, Wu, thanks the support of
Alexander von Humboldt Foundation.

\appendix
\section{Asymptotic Conditions on Horizon}

 In \cite{c2}, two asymptotic conditions had been given
\be \label{ac1}
\frac{\chi_K^a\chi^b_K}{\chi_K^2}\hat\dl g_{ab} \to 0
\ee
and
\be  \label{ac2}
\chi_K^at^b\hat\dl g_{ab} \to 0
\ee
as $\chi_K^2 \to 0$.  Because
\begin{eqnarray}
\label{cb1}
\frac{\chi_K^a\chi_K^b}{\chi_K^2}\hat\dl g_{ab}&=&\frac{\chi_K^a\chi_K^b}{\chi_K^2}{\cal L}_{\xi}g_{ab}\nonumber\\
&=&\frac{2\chi_K^a\chi_K^b}{\chi_K^2}\del_a(Tl_b+Rn_b)\nonumber\\
&=&\frac{2}{\chi_K^2}(l^a-\frac{\chi_K^2}{2}n^a)(l^b-\frac{\chi_K^2}{2}n^b)[(\del_aT)l_b+T\del_al_b+(\del_aR)n_b+R\del_an_b]\nonumber\\
&=&\frac{2}{\chi_K^2}[-(\bD R)+Rl^al^b\del_an_b 
+\frac{\chi_K^2}{2}(\Dl R)-\frac{\chi_K^2}{2}Rn^al^b\del_an_b\nonumber\\
&{ }&\qquad +\frac{\chi_K^2}{2}(\bD
T)-\frac{\chi_K^2}{2}Tl^an^b\del_al_b -\frac{\chi^4}{4}(\Dl
T)+\frac{\chi^4}{4}Tn^an^b\del_al_b
] \nonumber \\
&=&\frac{2}{\chi_K^2}(-\bD R+\kappa\frac{\rho}{|\chi_K^{\ }|}R)+\Dl
R+\bD T+\kappa\frac{\rho}{|\chi_K^{\ }|}T+O(\chi_K^2),
\end{eqnarray}
the condition (\ref{ac1}) becomes
\begin{eqnarray}
\label{nb1} [\frac{1}{\chi_K^2}(-\bD
R+\kappa\frac{\rho}{|\chi_K^{\ }|}R)+\Dl R+\bD
T+\kappa\frac{\rho}{|\chi_K^{\ }|}T]_H=0.
\end{eqnarray}
The condition (\ref{ac2}) is equivalent to
\be
\label{ac2'}
\chi_K^a m^b \hat\dl g_{ab} \to 0 \qquad \mbox{as $\chi_K^2 \to 0$}.
\ee
Due to
\begin{eqnarray}
\chi_K^am^b\hat\dl g_{ab}&=&(l^a-\frac{\chi_K^2}{2}n^a)m^b{\cal L}_{\xi}g_{ab}\nonumber\\
&=&(l^a-\frac{\chi_K^2}{2}n^a)m^b[(\del_aT)l_b+T\del_al_b+(\del_aR)n_b+R\del_an_b\nonumber\\
&{ }&+(\del_bT)l_a+T\del_bl_a+(\del_bR)n_a+R\del_bn_a]\nonumber\\
&=&Tl^am^b\del_al_b+Rl^am^b\del_an_b-m^a\del_aR+Rl^am^b\del_bn_a+O(\chi_K^2) \nonumber \\
&=&-\iota T -\dl R + (\bar\pi+\bar\alpha+\beta)R +O(\chi_K^2),
\end{eqnarray}
Eq.(\ref{ac2'}) requires
\be
\iota T \sim -\dl R \sim O(\chi_K^2).
\ee
Namely, $\iota \sim O(\chi_K^2)$.  Therefore, the asymptotic conditions
on the horizon specify a special class of the vector fields which has the
form of $\xi^a=Tl^a+Rn^a$ with $R\sim O(\chi_K^2)$.

\section{Some useful relations}

 In this appendix, some useful relations for the null tetrad are listed.
In the proof of the relations, the identities in Appendix A of \cite{c2} are used.

First,
\begin{eqnarray}
l^al^b\del_al_b&=&\frac{1}{2}\bD l^2=0, \label{A1-1} \\
n^al^b\del_al_b&=&\frac{1}{2}\Dl l^2=0, \label{A1-2} \\
l^an^b\del_an_b&=&\frac{1}{2}\bD n^2=0, \label{A1-3} \\
n^an^b\del_an_b&=&\frac{1}{2}\Dl n^2=0, \label{A1-4} \\
m^al^b\del_al_b&=&\frac{1}{2}\dl l^2=0, \label{A1-5} \\
m^an^b\del_an_b&=&\frac{1}{2}\dl n^2=0. \label{A1-6}
\end{eqnarray}
Because of Eq.(\ref{[ln]2}),
\begin{eqnarray}
l^al^b\del_an_b&=&l^bn^a\del_al_b-\kappa\frac{\rho}{|\chi_K^{\ }|}n_bl^b=
\kappa\frac{\rho}{|\chi_K^{\ }|},  \label{A2-1} \\
n^al^b\del_an_b&=&-n^an^b\del_al_b=-n^bl^a\del_an_b-\kappa\frac{\rho}{|\chi_K^{\ }|}n_bn^b=
0,   \label{A2-2}    \\
l^an^b\del_al_b&=&-l^al^b\del_an_b=-l^bn^a\del_al_b+\kappa\frac{\rho}{|\chi_K^{\ }|}n_bl^b=
-\kappa\frac{\rho}{|\chi_K^{\ }|}, \label{A2-3}  \\
n^an^b\del_al_b&=&n^bl^a\del_an_b+\kappa\frac{\rho}{|\chi_K^{\ }|}n_bn^b=0.  \label{A2-4}
\end{eqnarray}
In addition,
\begin{eqnarray} \label{A4-1}
\del_al^a &=&\frac{1}{2}(\del_a\chi_K^a+\tD \frac{|\chi_K^{\ }|}{\rho}+
\frac{|\chi_K^{\ }|}{\rho}\del_a\rho^a)\nonumber\\
&=&\frac{1}{2}\{-\frac{1}{2}\frac{\rho}{|\chi_K^{\ }|}\tD \frac{\chi_K^2}{\rho^2}
+\frac{|\chi_K^{\ }|}{\rho}[-2\kappa\frac{\rho^2}{\chi_K^2}+O(\chi_K^2)]\}\nonumber\\
&=&\kappa\frac{\rho}{|\chi_K^{\ }|}+O(\chi_K^2).
\end{eqnarray}
\begin{eqnarray} \label{A4-2}
\del_an^a
&=&-\frac{1}{\chi_K^2}\del_a\chi_K^a-\chi_K^a\del_a \frac{1}{\chi_K^2}+\frac{|\chi_K^{\ }|}{\rho}\tD
\frac{1}{\chi_K^2} + \frac{1}{\chi_K^2}\tD\frac{|\chi_K^{\ }|}{\rho}
+\frac{1}{\chi_K^2} \frac{|\chi_K^{\ }|}{\rho} \del_a\rho^a\nonumber\\
&=&\frac{|\chi_K^{\ }|}{\rho}\tD \frac{1}{\chi_K^2}
 +\frac{1}{\chi_K^2}\tD\frac{|\chi_K^{\ }|}{\rho}
+\frac{1}{\chi_K^2} \frac{|\chi_K^{\ }|}{\rho} \del_a\rho^a \nonumber\\
&=&\frac{|\chi_K^{\ }|}{\rho}\frac{1}{\chi^4}(2\kappa)\rho^2 -
\frac{1}{2}\frac{1}{\chi_K^2}\frac{\rho}{|\chi_K^{\ }|}\tD \frac{\chi_K^2}{\rho^2}
+\frac{1}{\chi_K^2}\frac{|\chi_K^{\ }|}{\rho}\del_a\rho^a\nonumber\\
&=&-2\kappa\frac{1}{\chi_K^2}\frac{\rho}{|\chi_K^{\ }|}-
\frac{1}{2}\frac{1}{\chi_K^2}\frac{\rho}{|\chi_K^{\ }|}O(\chi_K^2)
+\frac{1}{\chi_K^2}\frac{|\chi_K^{\ }|}{\rho}[-2\ka \d {\rho^2}{\chi_K^2}+O(\chi_K^2)]\nonumber\\
&=& O(1).
\end{eqnarray}


\begin{thebibliography}{99}
\bibitem{note1} It is more or less straightforward to formulate
the Noether charge formalism on an arbitrary
dimensional spacetime manifold.
\bibitem{note} In order to get rid of the confusion with the surface
gravity $\kappa$, we denote the N-P coefficient $-l^am^b \del_a l_b$
by $\iota$.
\end{thebibliography}
\end{document}